\documentclass[11pt,a4paper]{article}
\pdfoutput=1
\usepackage{jheppub}

\usepackage{amsmath}
\input epsf
\usepackage{epsfig}
\usepackage{amssymb}
\usepackage{graphics}
\usepackage[active]{srcltx}
\usepackage{epstopdf}

\newcommand{\bea}{\begin{eqnarray}}
\newcommand{\eea}{\end{eqnarray}}
\newcommand{\bean}{\begin{eqnarray*}}
\newcommand{\eean}{\end{eqnarray*}}
\newcommand{\nn}{\nonumber \\}

\def\W #1{\widetilde{#1}}
\def\WH #1{\widehat{#1}}

\def\braket#1{\left\langle #1 \right\rangle}

\def\ket#1{\left| #1\right\rangle}
\def\gb #1{ \left\langle #1 \right]}

\def\eref#1{(\ref{#1})}

\def\wt{\widetilde}

\def\a{{\alpha}}

\def\eps{\epsilon}

\def\vev{\braket}

\def\ket#1{\left| #1\right\rangle}
\def\bket#1{\left| #1\right]}

\def\Spaa{\vev}

\def\Spab{\gb}

\newcommand{\cA}{{\cal A}}

\newcommand{\cI}{{\cal I}}

\def\Label#1{\label{#1}%
  \smash{\hbox to0pt{\raise1ex\hbox{\tiny[#1]}\hss}}}

\def\oneloop{\tiny\mbox{1-loop}}

\def\tree{\tiny\mbox{tree}}
\def\cyclic#1{\mbox{Cyclic}\{#1\}}

\def\spaa #1{\langle #1\rangle}
\def\spbb #1{[#1]}
\def\spab #1{\langle #1]}

\title{The ${\cal Q}$-cut Representation of One-loop Integrands and Unitarity Cut Method}
\author[a]{Rijun Huang,}
\author[a]{Qingjun Jin,}
\author[a]{Junjie Rao,}
\author[a]{Kang Zhou}
\author[a,b]{and Bo Feng\footnote{The
unusual ordering of authors is just to let authors get proper
recognition of contributions under outdated practice in China. }}

\affiliation[a]{Zhejiang Institute of Modern Physics, Department of Physics,
 Zhejiang University, Hangzhou, 310027, P.R. China}
\affiliation[b]{Center of Mathematical Science,
  Zhejiang University, Hangzhou, 310027, P.R. China}

\emailAdd{huang@nbi.dk}
\emailAdd{qingjun@zju.edu.cn}
\emailAdd{raojunjie@zju.edu.cn}
\emailAdd{11236072@zju.edu.cn}
\emailAdd{fengbo@zju.edu.cn}

\date{\today}
\abstract{Recently, a new construction for complete loop integrands of massless field theories
has been proposed, with on-shell tree-level amplitudes delicately incorporated into its
algorithm. This new approach reinterprets integrands in a novel form, namely
the $\mathcal{Q}$-cut representation. In this paper, by deriving one-loop integrands as examples, we elaborate
in details the technique of this new representation, e.g., the summation over all possible $\mathcal{Q}$-cuts
as well as helicity states for the non-scalar internal particle in the loop.
Moreover, we show that the integrand in the ${\cal Q}$-cut representation naturally reduces
to the integrand in the traditional unitarity cut method for each given cut channel, providing a
cross-check for the new approach.
}

\begin{document}
\maketitle \flushbottom

%\newpage
%%%%%%%%%%%%%%%%%%%%
\section{Introduction}
\label{secIntroduction}
%%%%%%%%%%%%%%%%%%%

The attempt to calculate scattering amplitudes at tree and
loop-levels over the past decade (see reviews, e.g.,
\cite{Bern:2007dw,Elvang:2013cua,Henn:2014yza}) has revealed the
great computational ability of on-shell methods. For example, the
recursive method, such as the BCFW recursion relation
\cite{Britto:2004ap,Britto:2005fq}, enables one to build all
$n$-point tree-level amplitudes purely through the simplest on-shell
objects, e.g., the 3-point tree amplitudes, for a broad range of
field theories\footnote{For some other theories, the boundary
contribution eventually shows up under the familiar BCFW deformation
\cite{ArkaniHamed:2008yf,Cheung:2008dn}. Recently, several works had
been devoted to understanding the boundary contribution
\cite{Jin:2014qya,Jin:2015pua,Feng:2014pia,Feng:2015qna}.
Alternative methods can also be found in
\cite{Cheung:2015cba,Cheung:2015ota}.}. Another powerful on-shell
technique is the unitarity cut \cite{Bern:1994zx,Bern:1994cg} (with
its generalization to $D$-dimension
\cite{Anastasiou:2006jv,Anastasiou:2006gt}), or the generalized
unitarity method \cite{Britto:2004nc,Britto:2005ha} which attacks
amplitudes at loop-level. Combined with the reduction methods (for
example, the computational algebraic geometry methods
\cite{Zhang:2012ce,Mastrolia:2012an}), it greatly simplifies the
construction of multi-loop amplitudes, through lower-loop ones or
even solely tree amplitudes.

The success of these on-shell methods, especially the BCFW recursion
relation at tree-level, inspires  people to search for an on-shell
construction for loop amplitudes
\cite{CaronHuot:2010zt,ArkaniHamed:2010kv,Boels:2010nw}. However,
for the latter there are more subtle issues. First of all, when
talking about loop amplitudes, we should clarify two terminologies,
namely the {\sl loop integrand} and  the {\sl loop integral after
integration}. Due to the fact that, the integrand is a rational
function of loop as well as external momenta, whose analytic
behavior is very similar to tree amplitudes, it is natural to expect
a similar recursion relation for it. Although such intuition is
correct, the rational function at loop-level possesses two major
differences, compared to that of tree-level, which might greatly
obstruct the construction of loop integrands. The first difference
is the ambiguity of how to define the loop momentum. Under proper
regularizations \cite{tHooft:1972fi,Collins:1986,Bern:1991aq} (such
as the dimensional regularization), we can arbitrarily shift the
loop momentum while keeping its integration invariant. Because of
this degree of freedom, in general it is difficult to identify a
unique definition simultaneously for all Feynman diagrams (the
planar diagram is an exception, where a canonical ordering of loop
momenta do exist thanks to color ordering
\cite{ArkaniHamed:2010kv}). The second crucial difference is the
so-called forward limit. When reducing $n$-point $L$-loop Feynman
diagrams to $(n+2)$-point $(L-1)$-loop ones by setting an internal
propagator on-shell
%\footnote{or "cut" a propagator, which corresponds to taking the residue of the pole coming from internal propagator under ordinary BCFW-deformation of two external momenta.}
, we do not get
the complete set of $(n+2)$-point $(L-1)$-loop Feynman diagrams,
since those ones containing forward singularities have been
automatically excluded. Thus, when constructing $L$-loop
integrands recursively from the complete $(L-1)$-loop integrands, we
should manually exclude those singular parts, and this is not
an easy task. Fortunately for some theories, the forward
singularities do not exist. One nice example is the planar $\mathcal{N}=4$ super-Yang-Mills (SYM)
theory, for which the all-loop recursion relation
has been written down, thanks to the absence of
two difficulties mentioned above \cite{ArkaniHamed:2010kv}\footnote{There are
also some discussions regarding the single cut method, for which the forward limit problem can be (partly)
avoided \cite{NigelGlover:2008ur,Bierenbaum:2010cy,Elvang:2011ub,Britto:2010um}.}.

%Although fortunately for some theories, the forward singularities do not exist and the forward limit trouble can be (or partly) avoided\cite{NigelGlover:2008ur,Bierenbaum:2010cy,Elvang:2011ub,Britto:2010um,CaronHuot:2010zt,ArkaniHamed:2010kv}.
%An convincing example is the $\mathcal{N}=4$ super-Yang-Mills(SYM) theory, where recursion relation for all-loop-level of planar diagram part has been written down, thanks to the absence of above mentioned two difficulties\cite{ArkaniHamed:2010kv}.

When considering generic theories, the two difficulties
mentioned above are inevitable, thus a new idea to handle them is in
demand. Quite recently, the authors of ref. \cite{Baadsgaard:2015twa}
provided a new algorithm, namely the $\mathcal{Q}$-cut
construction, which delicately resolves these issues\footnote{At this
moment, the ${\cal Q}$-cut construction is restricted to massless
theories. So throughout this paper, we will only focus on
massless theories.}. This algorithm derives an expression (namely
the $\mathcal{Q}$-cut representation) of loop integrands which looks
quite different from the familiar one generated from Feynman
diagrams. Here, a canonical way of defining loop
momenta can be prescribed, regardless of the loop amplitude considered
is color-ordered or not. Also, with a certain kind of scale
deformation, the forward singularities can be neatly stripped off.

Although the general framework of $\mathcal{Q}$-cut construction has
been settled in \cite{Baadsgaard:2015twa}, in order to
completely understand its structure, further demonstrations in details
are still needed. For instance, when summing over all possible
$\mathcal{Q}$-cut terms, what rules shall we follow to avoid
over-counting or missing some terms? Furthermore, if the internal
particle of the loop is not scalar but fermion or gluon, how
shall we sum over the helicity states? Besides, the
divergence after the loop integration demands us to
properly regularize it. Since the most common
regularization scheme is the dimensional regularization, we would
like to see how ${\cal Q}$-cut construction fits itself into this scheme.
Addressing these points with ample examples of one-loop
amplitudes would be our first pursuit in this paper.

We want to emphasize that, the ${\cal Q}$-cut construction is a {\sl
complete} algorithm. It gives the construction of loop
integrands in ${\cal Q}$-cut representation {\sl and} the correct
physical contours to do the loop integrations \cite{Baadsgaard:2015twa}.
%integration of integral by contour integration described in, finally leads to the physical loop amplitude.
However, since the ${\cal Q}$-cut representation is very different
from the familiar integrand generated by Feynman diagrams with
quadratic propagators, it would be illustrative to perform a
cross-check by using another completely independent algorithm. Our second
pursuit here is to provide such a cross-check via the
unitarity cut method. We will show that, for each unitarity cut of a given
one-loop amplitude, the contribution from ${\cal Q}$-cut
representation is identical to the product of two on-shell tree
amplitudes under the traditional unitarity cut, thus
verifying the equivalence between these two algorithms.

This paper is organized as follows. In \S \ref{secReview}, we will
review the $\mathcal{Q}$-cut representation and unitarity cut
method, and discuss the connection between them. In \S
\ref{secScalar} and \S \ref{secYangMills}, we use various examples
of one-loop amplitudes in scalar field and Yang-Mills theories to
demonstrate the details of $\mathcal{Q}$-cut representation, and
perform a cross-check by the unitarity cut. The conclusion and
discussions are given in \S \ref{secConclusion}, and in the
appendix, conventions for helicity choice of gluon in $D$-dimension
as well as all $D$-dimensional 4-point tree amplitudes are given.

%the momentum configuration of  $(n+2)$ external legs is not general, but the forward limit where momenta of two external legs go to $\ell,-\ell$.

%topology into a lower connected tree topology (for example, cutting one internal propagator to open one loop diagram to a tree), the obtained tree diagrams do not correspond to the full tree level amplitude at the forward limit (remembering that after the cut, each  internal line becomes two extra external legs with momenta $\ell, -\ell$), since in general the full tree level amplitude will contain these Feynman diagrams where external legs with momenta $\ell, -\ell$ attached to the cubic vertex. The exclusion of these forward singularities from the full

%%%%%%%%%%%%%%%%%%%%%
\section{Review and general discussions}
\label{secReview}
%%%%%%%%%%%%%%%%%%%%%%%%

In this section, we will firstly review the new
$\mathcal{Q}$-cut construction of loop integrands proposed in
\cite{Baadsgaard:2015twa}. Since its result is quite different from
the one given by standard Feynman diagrams (where the propagators
are quadratic), understanding this new structure from
various aspects would be of interest. As mentioned in the
introduction, we will provide such an digestion based on the
unitarity cut method. So the relevant background of unitarity cut
will be reviewed shortly afterwards. Then we will present the
general aspects of how these two different algorithms are connected.

%%%%%%%%%%%%%%%%%%%%%%%
\subsection{Construction of the ${\cal Q}$-cut representation}
%%%%%%%%%%%%%%%%%%%%%%

The working experience of on-shell recursion relations at tree-level
tells us that, a nice way of determining a rational function is to
apply the residue theorem by some proper deformation. For tree
amplitudes, the BCFW deformation is the simplest choice involving
minimal number of external legs, while maintaining the on-shell
condition and momentum conservation. For loop integrands, similar
attempt is not successful due to the two difficulties mentioned
before. However, the ambiguity of defining the loop momentum, on the
other hand, also allows us to shift its components. Thinking it
further, to avoid its entanglement with external momenta, we could
shift its components in extra dimensions. For loop momenta, this
operation is feasible, since in the dimensional regularization
scheme, although all external momenta are kept in 4-dimension, we do
need to take loop momenta into $(4-2\eps)$-dimension, i.e., $\W
\ell_{4-2\eps}=(\ell_{{\tiny \mbox{4-dim}}}, \vec{\mu}_{-2\eps})$.
With this observation, we can shift the loop momentum as $\W\ell\to
\W\ell+\vec{\eta}$. So the internal propagators are deformed as
$(\W\ell+P)^2 \to (\W\ell+P)^2+z$ ($z=\vec{\eta}^2$) with arbitrary
4-dimensional momentum $P$, provided $\W\ell\cdot \vec{\eta}=0$.
Such a shift can be achieved by either shifting $\vec{\eta}$ in
extra dimensions (so $\W\ell_{4-2\eps}\to
(\W\ell_{4-2\eps},\vec{\eta})$), or keeping $\vec{\eta}$ in the
$(-2\eps)$-dimension (so $\W\ell_{4-2\eps}\to (\ell_{{\tiny
\mbox{4-dim}}},\vec{\mu}+\vec{\eta})$) with additional condition
$\vec{\mu}\cdot \vec{\eta}=0$. No matter which way is taken, the
shift is always performed in extra dimensions. Thus we may call it
the {\sl dimensional deformation}. To avoid confusion, hereafter we
will use $\W\ell$ to denote the loop momentum in
$(4-2\eps)$-dimension and $\ell$ the 4-dimensional component, i.e.,
$\W \ell=(\ell, \vec{\mu})$. Moreover, we will use $\WH \ell=\W\ell
+\vec{\eta}$ to denote the deformed loop momentum.

Then, we can continue to discuss the rational function
${\cal I}(\W\ell)$ obtained from Feynman diagrams, and consider the
familiar contour integration $\oint {dz\over z} {\cal I}(\WH \ell)$.
Unlike the tree amplitude, for this case it is easy to check that
its boundary contribution (i.e., the residue at $z=\infty$) is a scale-free
rational function in terms of $\W\ell$. Thus it integrates to zero under
dimensional regularization and hence can be dropped. The residues of
finite poles take the form
$${1\over
(\W\ell+P_0)^2} \left\{ { {\cal N}(\WH\ell) \over \prod (\WH\ell+
P_i)^2}\right\}_{(\WH\ell+P_0)^2=0}~,$$ where the condition
$(\WH\ell+P_0)^2=0$ means nothing but putting this internal
propagator on-shell (in higher dimension). More specifically, while
${1\over (\W\ell+P_0)^2}$ is an off-shell internal propagator, the
expression inside the bracket is an on-shell tree amplitude. Such a
form has exactly the same structure as the tree-level BCFW formula
${1\over P^2} \left\{ A_L(\WH P) A_R(-\WH P) \right\}$. Once the
residues of all finite poles are obtained, we can perform a further
shift term-by-term by translating $\W\ell+P_0 \to \W\ell$, such that
each off-shell propagator ${1\over (\W\ell+P_0)^2}$ is replaced by
${1\over \W\ell^2}$. The legal shift of loop momenta under proper
regularization will not alter the final result after integration,
and this kind of shift leads us to the canonical definition of loop
momentum\footnote{The canonical way of defining the loop momentum is
given in  \cite{Geyer:2015bja}.}.

\begin{figure}
\centering
  \includegraphics[width=5.5in]{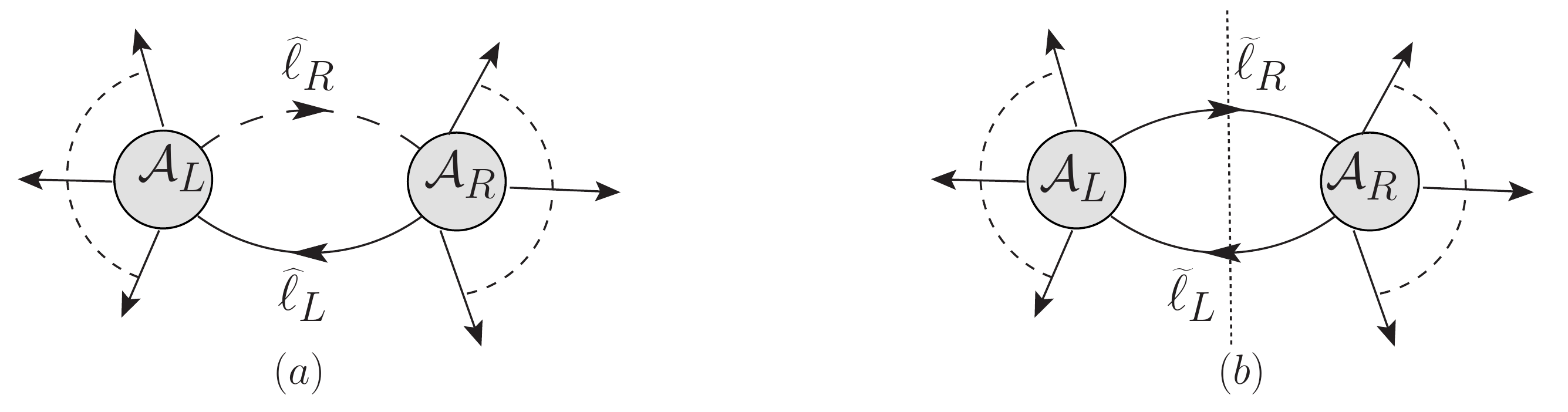}\\
  \caption{(a) Graphic presentation of ${\cal Q}$-cut: the tree amplitudes are evaluated with the
  rescaled $D$-dimensional loop momenta $\widehat{\ell}_L$ and $\widehat{\ell}_R$, multiplied
  by two novel propagators $1/\ell^2$ and $1/(2\ell\cdot P_L+P_L^2)$. (b) Graphic presentation
  of unitarity cut: the tree amplitudes are evaluated with the on-shell
  loop momenta $\widetilde{\ell}_L$ and $\widetilde{\ell}_R$,
  with two propagators $1/\widetilde{\ell}_L^2\widetilde{\ell}_R^2$ replaced by
  $\delta^{+}(\widetilde{\ell}_L^2)\delta^{+}(\widetilde{\ell}_R^2)$.}\label{QUcut}
\end{figure}

Next, we need to impose on-shell conditions for the
expression inside the bracket. This is achieved by rewriting quadratic
propagator $(\WH \ell+P)^2$ as a linear propagator $(2\W\ell\cdot
P+P^2)$. More specifically, under the dimensional deformation, we will
arrive at expressions like
\bea {\cal I}^{\cal Q}_{{\tiny\mbox{step-1}}}(\W\ell)= \sum {1\over
\W\ell^2} \left[{{\cal N}(\W\ell)\over \prod (2\W\ell\cdot
P_i+P_i^2)}\right]~.~~~\label{Q-cut-step-1}\eea
As emphasized in ref. \cite{Baadsgaard:2015twa}, the forward limit
singularities prevent us from interpreting ${\cal I}^{\cal Q}_{{\tiny
\mbox{step-1}}}$ as the full on-shell tree amplitudes. In
order to obtain a well-defined result, we need to take a second
deformation, namely the {\sl scale deformation}\footnote{It is worth to notice that
the scale deformation will keep the null momentum $\W\ell$ to be null. } $\W\ell\to \a
\W\ell$ and then evaluate the contour integration $\oint {d\a\over
(\a-1)} {\cal I}^{\cal Q}_{{\tiny \mbox{step-1}}}(\a\W\ell)$. After
dropping the residues\footnote{As stated in
\cite{Baadsgaard:2015twa}, these residues precisely correspond to
the ill-defined terms in the forward limit.} of poles at $\a=0$
and $\a=\infty$, which are scale-free terms so they integrate to zero,
we finally arrive at the {\sl ${\cal Q}$-cut
representation}\footnote{We will call the algorithm above with two-step
deformation as the {\sl ${\cal Q}$-cut construction}.} of the loop integrand,
\bea \cI_{n}^{\cal Q}(\ell)=\sum_{P_L}\sum_{h_1,h_2}\cA_{L}(\cdots,
\widehat{\ell}_R^{~h_1},-\widehat{\ell}_L^{~h_2}){1\over
\W\ell^2}{1\over (-2\W\ell\cdot
P_L+P_L^2)}\cA_R(\widehat{\ell}_L^{~\bar{h}_2},-\widehat{\ell}_R^{~\bar{h}_1},\cdots)~,
~~~\label{q-cut}\eea
where $\widehat{\ell}_L=\alpha_L(\W\ell+\vec{\eta})$,
$\widehat{\ell}_R\equiv \widehat{\ell}_L-P_L$ with
$\alpha_L=P_L^2/(2\W\ell\cdot P_L)\neq 0$ and
$\vec{\eta}^2=\W\ell^2$ (see Figure (\ref{QUcut}.a)).

Let us explain more details of the ${\cal Q}$-cut representation
\eref{q-cut}. Firstly, although $\vec{\eta}$ does appear in
$\widehat{\ell}_L$, it should be eliminated together with $\cA_L, \cA_R$
by the on-shell condition $\vec{\eta}^2=\W\ell^2$. After that, all
propagators involving $\WH \ell$ become linear in $\W\ell$.
Secondly, since in this algorithm two different deformations have been
performed and each one sets an internal
propagator on-shell, these two on-shell propagators are not
equivalent. This is entirely different from unitarity
cut where the two cut propagators are essentially equivalent. Due to
this difference, the sum over $P_L$ (with condition $P_L^2\neq
0$) includes all possible partitions of external legs associated with
${\cal Q}$-cut, {\sl without} modulo the cyclic group. Thus, as we will see soon, two
terms in ${\cal Q}$-cut representation of momenta $P_L$ and
$-P_L$ correspond to a single unitarity cut of momentum
$P_L$\footnote{Taking the 4-point one-loop amplitude as an example,
both choices $P_L=p_1+p_2$ and $P_L=p_3+p_4$ correspond to
the $s_{12}$ channel in unitarity cut, but they
should be regarded as two different $\mathcal{Q}$-cuts and summed
over to reach the full integrand.}.

Now we move to the second summation in \eref{q-cut}, which
should be carefully treated when the internal particle along
$\ell_L$ or $\ell_R$ is not a scalar. As emphasized before, in the
${\cal Q}$-cut construction, we have met three different kind of dimensions:
the plain 4-dimension for external momenta, the
$(4-2\eps)$-dimension for loop momenta under the dimensional
regularization, and the $(4-2\eps+d)$-dimension for the deformed
loop momenta. A question naturally arises: when we sum over helicity
states of internal particles, which dimension should we use, the
$(4-2\eps)$-dimension or the $(4-2\eps+d)$-dimension? The correct
treatment is to sum over helicity states in the
$(4-2\eps)$-dimension. This can be explained via the following two
arguments. Firstly, according to the Feynman rules in dimensional
regularization, before setting the internal propagator on-shell, the
metric $\eta_{\mu\nu}$ we use is of $(4-2\eps)$-dimension. Secondly,
as explained before, the dimensional deformation can be also
interpreted as deforming $\vec{\mu}\to \vec{\mu}+\vec{\eta}$ in the
$(-2\eps)$-dimension with additional condition $\vec{\mu}\cdot
\vec{\eta}=0$, in which case we have not
defined the $(4-2\eps+d)$-dimension at all.

Before concluding this part, we will give a brief remark. While
enjoying all advantages of on-shell tree amplitudes as building
blocks, the ${\cal Q}$-cut representation systematically produces a
completely off-shell one-loop integrand. This is different from the
traditional unitarity cut method, where after getting all pieces of
integrand for each unitarity cut, we should assemble them carefully
to produce the full integrand, avoiding over-counting or missing
terms. The only price (or novelty) of these accomplishments in the
new approach is to replace the quadratic propagators by rescaled
linear propagators.

%%%%%%%%%%%%%%%%%
\subsection{Unitarity cut method}
%%%%%%%%%%%%%%%%%

The Passarino-Veltman reduction \cite{Passarino:1978jh} is a standard
method of computing loop amplitudes. In its context, an one-loop
amplitude $\cA$ of massless theories can be expanded as some master
integrals \cite{Bern:1992em,Bern:1993kr} as $\cA=\sum_i c_i \int
d^{4-2\eps} \ell ~\cI_i$, where $c_i$ is an expansion
coefficient (a rational function of external kinematic data) and $\cI_i$ is
a scalar integrand of pentagon, box, triangle and bubble
topologies. Thus the computation of generic one-loop amplitudes can
be reduced to determining of coefficients $c_i$, while
the unitarity cut method is good for this
purpose \cite{Anastasiou:2006jv,Anastasiou:2006gt,Britto:2004nc,Britto:2005ha,Britto:2006sj,Ossola:2006us,
Britto:2006fc,
Britto:2007tt,Forde:2007mi,Ellis:2007br,Giele:2008ve,Ossola:2008xq,Badger:2008cm}.

More explicitly, for the unitarity cut in $s_P$ channel
with respect to the cut momentum $P$, as shown in Figure
(\ref{QUcut}.b), we evaluate
\bea \Delta \cA^{\oneloop}_n(\widetilde{\ell})|_{P}=\int d\Omega
~\cI^{\oneloop}_n(\widetilde{\ell})|_{P}~,~~~\label{Uni}\eea
where $\cI^{\oneloop}_n(\widetilde{\ell})|_{P}\equiv\cI
~\widetilde{\ell}_L^2~\widetilde{\ell}_R^2$ is the cut integrand
obtained via multiplying the full one-loop integrand by two cut
propagators $\widetilde{\ell}_L^2,\widetilde{\ell}_R^2$. The
integration measure above is given by
\bea \int d\Omega=\int d^{\dim[\wt\ell]}\widetilde{\ell}_L~
d^{\dim[\wt\ell]}\widetilde{\ell}_R~\delta^+(\widetilde{\ell}_L^{~2})\delta^+(\widetilde{\ell}_R^{~2})
\delta^{\dim[\wt\ell]}(\widetilde{\ell}_R-\widetilde{\ell}_L+P)~.~~~\label{Omega}
\eea
Here, $\Delta \cA^{\oneloop}_n(\widetilde{\ell})|_{P}$ is the imaginary
part with respect to $P$. It is also crucial to
note that in unitarity cut, cut momentum $P$ is equivalent
to $-P$, since it just corresponds to $\widetilde{\ell}_R
\leftrightarrow \widetilde{\ell}_L$. Thus when talking about all
possible cuts, we should in fact consider all inequivalent $\{P,-P\}$ pairs.

The unitarity cut can be applied to the one-loop amplitude
reduction. Since $c_i$ has no branch cut, we have
$\Delta\cA=\sum_i c_i \Delta\int d^{4-2\eps} \W\ell ~\cI_i$.
Furthermore, for different master integrals, $\Delta\int d^{4-2\eps}
\W\ell ~\cI_i$ are distinct analytic functions. Thus by comparing
both sides of the expansion, we can determine coefficients $c_i$.

The central idea of the unitarity cut method relies on the fact
that, the cut integrand $\cI^{\oneloop}_n(\widetilde{\ell})|_{P}$ is
given by
\bea
\cI^{\oneloop}_n(\widetilde{\ell})|_{P}=\sum_{h_1,h_2}\cA_{L}(\cdots,
\widetilde{\ell}_R^{~h_1},-\widetilde{\ell}_L^{~h_2})\cA_R(\widetilde{\ell}_L^{~\bar{h}_2},
-\widetilde{\ell}_R^{~\bar{h}_1},\cdots)~.~~~\label{u-cut}\eea
Thus when computing the expansion coefficients, we do not have to
start with the full one-loop integrand, which is usually given by
Feynman diagrams. Since the on-shell tree amplitudes are much
simpler to compute, the unitarity cut indeed greatly improves the
efficiency of loop amplitude computation.

%%%%%%%%%%%%%%%%%%%%
\subsection{Connection between two approaches}
%%%%%%%%%%%%%%%%%%%%

Having reviewed the ${\cal Q}$-cut construction and the unitarity
cut method, let us move to one of our major concerns, i.e., understanding the
${\cal Q}$-cut construction. Since this is a complete approach, we should have
\bea \int d^{\dim[\wt\ell]}\widetilde{\ell}~\cI^{\cal
Q}(\widetilde{\ell})=\int
d^{\dim[\wt\ell]}\widetilde{\ell}~\cI^{\cal
F}(\widetilde{\ell})~,~~~\label{purpose}\eea
where $\cI^{\cal Q}$ and $\cI^{\cal F}$ are loop integrands produced
by ${\cal Q}$-cut construction and Feynman diagrams respectively.
For generic theories, the loop integrand of course will be very
complicated and these two integrands would look quite different. Thus
their direct comparison is practically impossible for most of
cases, let alone further integrating them out. However, as we have
mentioned, the calculation of one-loop amplitudes is equivalent to
fixing all coefficients $c_i$ of master integrals. If we can show
that, the same coefficients can be obtained by using
$\cI^{\cal Q}(\widetilde{\ell})$ as the input for \eref{Uni}, the
identity \eref{purpose} must hold.

Let us consider the following expression
\bea & &\int d^{\dim[\wt\ell]}\widetilde{\ell}_L~
d^{\dim[\wt\ell]}\widetilde{\ell}_R~\delta^+(\widetilde{\ell}_L^{~2})\delta^+(\widetilde{\ell}_R^{~2})
\delta^{\dim[\wt\ell]}(\widetilde{\ell}_R-\widetilde{\ell}_L+P)\nonumber\\
&&~~~~~~~~~~~~~~~~~~~~~~~~~~\times\left\{\widetilde{\ell}_L^{~2}~\widetilde{\ell}_R^{~2}
\sum_{P_L}\sum_{h_1,h_2}{\cA_{L}(\cdots,
\widehat{\ell}_R^{~h_1},-\widehat{\ell}_L^{~h_2})
\cA_R(\widehat{\ell}_L^{~\bar{h}_2},-\widehat{\ell}_R^{~\bar{h}_1},\cdots)\over
\W\ell^2 (-2\W\ell\cdot P_L+P_L^2)}\right\}~,~~~\label{Qcut-uni}\eea
where the explicit expression of $\cI^{\cal Q}(\widetilde{\ell})$
in \eref{q-cut} has been inserted. In the traditional Feynman
diagram approach, it is simple to identify $\widetilde{\ell}_L$ and
$\widetilde{\ell}_R$ with the corresponding propagators in $\cI^{\cal
F}(\widetilde{\ell})$. However, in the ${\cal Q}$-cut
representation, under the canonical definition of loop momentum for each
$\mathcal{Q}$-cut term, there is one and only one quadratic
propagator in $\cI^{\cal Q}(\widetilde{\ell})$. Hence two possible
identifications are allowed, one is
$\widetilde{\ell}=\widetilde{\ell}_L$ and the other
$\widetilde{\ell}=\widetilde{\ell}_R$. Thanks to the factor
$\delta^+(\widetilde{\ell}_L^2)\delta^+(\widetilde{\ell}_R^2)\widetilde{\ell}_L^2~\widetilde{\ell}_R^2$,
it is easy to see that when taking
$\widetilde{\ell}=\widetilde{\ell}_L$, the only surviving term is
the one with $P_L=P$, and expression \eref{Qcut-uni} reduces to
\bea & &\int d^{\dim[\wt\ell]}\widetilde{\ell}_L~
d^{\dim[\wt\ell]}\widetilde{\ell}_R~\delta^+(\widetilde{\ell}_L^{~2})\delta^+(\widetilde{\ell}_R^{~2})
\delta^{\dim[\wt\ell]}(\widetilde{\ell}_R-\widetilde{\ell}_L+P)
\sum_{h_1,h_2}\cA_{L}(\cdots,
\widehat{\ell}_R^{~h_1},-\widehat{\ell}_L^{~h_2})
\cA_R(\widehat{\ell}_L^{~\bar{h}_2},-\widehat{\ell}_R^{~\bar{h}_1},\cdots)~.~~~\label{Qcut-uni-I}\eea
Recalling $\widehat{\ell}_L=\alpha_L(\W\ell+\vec{\eta})$,
$\widehat{\ell}_R\equiv \widehat{\ell}-P_L$ with
$\alpha_L=P_L^2/(2\W\ell\cdot P_L)\neq 0$ and
$\vec{\eta}^2=\W\ell^2$, by using $\W\ell^2=\W\ell_L^2=0$ we
obtain $\vec{\eta}=0$. Furthermore, with
$\W\ell_R^2=(\W\ell-P_L)^2=0$ we obtain $P_L^2=(2\W\ell\cdot P_L)$,
thus $\a_L=1$. Putting all pieces together, we find
$\widehat{\ell}_L=\W\ell_L$ and $\widehat{\ell}_R=\W\ell_R$ in the
given unitarity cut, then expression \eref{Qcut-uni-I} reduces to
\bea & &\int d^{\dim[\wt\ell]}\widetilde{\ell}_L~
d^{\dim[\wt\ell]}\widetilde{\ell}_R~\delta^+(\widetilde{\ell}_L^{~2})\delta^+(\widetilde{\ell}_R^{~2})
\delta^{\dim[\wt\ell]}(\widetilde{\ell}_R-\widetilde{\ell}_L+P)
\sum_{h_1,h_2}\cA_{L}(\cdots, \W{\ell}_R^{~h_1},-\W{\ell}_L^{~h_2})
\cA_R(\W{\ell}_L^{~\bar{h}_2},-\W{\ell}_R^{~\bar{h}_1},\cdots)~.~~~\label{Qcut-uni-I-1}\eea
The expression above is exactly the one of \eref{Uni} with input
\eref{u-cut}. It tells that, taking ${\cal Q}$-cut representation
as the input, we can reproduce the same expansion coefficients $c_i$ as
those computed by the traditional unitarity cut method.

If we identify $\widetilde{\ell}=\widetilde{\ell}_R$,
the only surviving term in \eref{Qcut-uni} would be the one with
$P_L=-P$. After a analogous analysis, it can be checked that now
$\widehat{\ell}_L=\W\ell_R$ and $\widehat{\ell}_R=\W\ell_L$. Then
\eref{Qcut-uni} reduces to \eref{Qcut-uni-I-1} with the relabeling
$\cA_L\leftrightarrow \cA_R$, and it also produces the same
expansion coefficients $c_i$. In the subsequent part of this paper, we will
demonstrate the calculations above with explicit examples.

Now, we give a brief summary of these general arguments. By comparing
the expansion coefficients of master integrals produced by the
unitarity cut method, we have checked that the ${\cal Q}$-cut
representation indeed produces the complete one-loop integrand.
Furthermore, we see that while it is non-trivial to assemble the
results from all possible unitarity cuts into the full loop integrand
with some proper off-shell continuation, the ${\cal Q}$-cut
representation provides a more natural solution.

In the following sections, we will go through several examples and
clarify the details discussed above.

%%%%%%%%%%%%%%%%
\section{Applications in scalar field theory}
\label{secScalar}
%%%%%%%%%%%%%%%%%

The amplitudes of scalar field theory are simple enough to explore the
details of $\mathcal{Q}$-cut representation at the level of the full
integrand, while the pole structure of non-color-ordered
amplitudes resembles that of non-planar amplitudes. Hence it is
worthwhile to go through a few examples of both color-ordered and
non-color-ordered scalar amplitudes. Since the internal particles
of the loop are scalars, in this section, we will use
$\ell$ to denote the $(4-2\epsilon)$-dimensional $\widetilde{\ell}$ for simplicity.

%%%%%%%%%%%%%%%%%%%
\subsection{Warmup examples}
%%%%%%%%%%%%%%%%%%%

\begin{figure}
    \centering
  \includegraphics[width=5.5in]{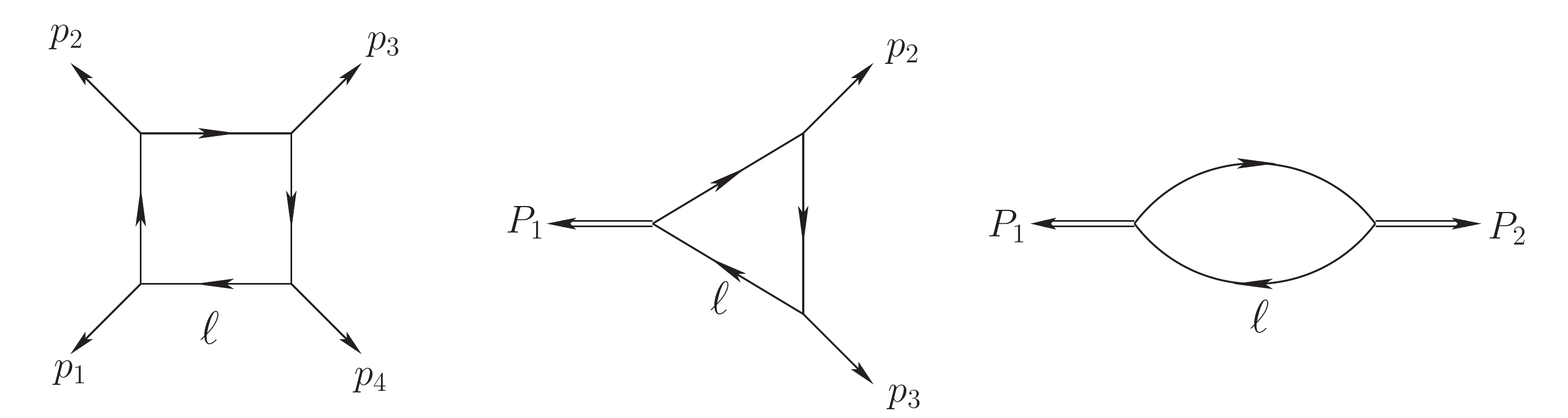}\\
  \caption{Convention of external momenta for scalar integrands of box, triangle and
   bubble topologies. $p_i$ denotes massless momentum and $P_i$ denotes the sum of several
   massless momenta.}\label{Pintegrals}
\end{figure}

Before heading to explicit amplitudes, let us start with some warmup
exercises, i.e., the scalar integrands\footnote{The integrands whose
numerator is 1 and denominator is product of propagators.} of
bubble, triangle and box topologies, illustrated in Figure
(\ref{Pintegrals}). With these examples, we want to clarify the
differences between the results of $\mathcal{Q}$-cut construction
and the original scalar integrands. These results will be useful for
the later discussion of explicit amplitudes.

~\\
{\bf Scalar integrand of bubble topology:}  Let us focus on the
expression
\bea {\cal I}^{\flat}_{\between}(P_1;P_2)={1\over
\ell^2(\ell-P_1)^2}~,~~~\label{F-bubble}\eea
where $P_1+P_2=0$, and $P_i$ is the sum of several massless momenta.
Applying the general partial fraction identity\footnote{The
importance of this identity has been demonstrated in
\cite{Geyer:2015bja}.},
\bea {1\over D_1\cdots D_m}=\sum_{i=1}^{m}{1\over
D_i}\left[\prod_{j\neq i}{1\over D_j-D_i}\right]~,~~~\label{PF}\eea
it becomes
\bea  {\cal I}^{\flat}_{\between}(P_1;P_2)&=&{1\over
\ell^2((\ell-P_1)^2-\ell^2)}+{1\over(\ell-P_1)^2(\ell^2-(\ell-P_1)^2)}~.~~~\eea
Since this result should be combined with the loop integration
$\int d^4\ell$, with proper regularization (such as dimensional
regularization), we can shift the loop momentum term-by-term without
altering the integrated result. Thus we can write
\bea {\cal I}^{\flat,pf}_{\between}(P_1;P_2)\simeq {1\over
\ell^2(-2\ell\cdot P_1+P_1^2)}+{1\over \ell^2(-2\ell\cdot
P_2+P_2^2)}~.~~~\label{bubbleQ} \eea
The symbol $\simeq$ means the two expressions are equivalent upon
integration. The superscript $^{pf}$ denotes the result after
partial fraction identity and momentum shifting.

We can also derive the ${\cal Q}$-cut representation of expression
\eref{F-bubble} by using the two-step deformation reviewed in the
previous section. For this simple case, the residues at finite poles
of $\alpha$ yields exactly ${\cal I}^{\flat,{\cal
Q}}_{\between}(P_1;P_2)= {\cal I}^{\flat,pf}_{\between}(P_1;P_2)$,
where the superscript ${\cal Q}$ denotes the result produced by
${\cal Q}$-cut construction. Hence upon integration, we have the
following two equivalent expressions
\bea {1\over \ell^2(\ell-P_1)^2}\simeq {1\over \ell^2(-2\ell\cdot
P_1+P_1^2)}+{1\over \ell^2(-2\ell\cdot P_2+P_2^2)}~,~~~\nonumber\eea
which can be used in later comparison.

~\\
{\bf Scalar integrand of triangle topology:} Let us focus on the
expression
\bea {\cal I}^{\flat}_{\triangle}(P_1;p_2;p_3)={1\over
\ell^2(\ell-P_1)^2(\ell+p_3)^2}~,~~~\label{F-tri}\eea
where again $P_1$ is the sum of several massless momenta. Applying
the partial fraction identity \eref{PF}, and then shifting the loop
momentum, we get
\bea {\cal I}^{\flat,pf}_{\triangle} &\simeq &{1\over
\ell^2(-2\ell\cdot P_1+P_1^2)(2\ell\cdot p_3)}+{1\over
\ell^2(2\ell\cdot P_1+P_1^2)(-2\ell\cdot p_2)}+{1\over
\ell^2(-2\ell\cdot p_3)(2\ell\cdot p_2)}~.~~~\label{F-tri-F-rep}\eea
The last term is scale-free and can be dropped when performing the
loop integration.

Now, we take a second deformation $\ell\to \alpha\ell$ and compute
the residues of ${{\cal I}^{\flat,pf}_{\triangle}(\alpha\ell)\over
\alpha-1}$\footnote{Remind that the overall ${1\over \ell^2}$ factor
is not shifted.} at finite poles of $\alpha$ excluding $\alpha=0,1$.
After dropping the scale-free terms, we get
\bea {\cal I}^{\flat,{\cal Q}}_{\triangle}(P_1;p_2;p_3)={\ell\cdot
P_1\over\ell^2(-2\ell\cdot P_1+P_1^2)P_1^2(\ell\cdot
p_3)}-{\ell\cdot p_{23}\over \ell^2(-2\ell\cdot p_{23}+p_{23}^2)
p_{23}^2(\ell\cdot p_2)}~.~~~\label{triangleQ} \eea
Unlike the bubble case, it is obvious that ${\cal I}^{\flat,{\cal
Q}}_{\triangle}(P_1;p_2;p_3)\neq {\cal
I}^{\flat,pf}_{\triangle}(P_1;p_2;p_3)$. Nonetheless, it is yet easy
to figure out that
\bea  {\cal I}^{\flat,{\cal Q}}_{\triangle}(P_1;p_2;p_3)-{\cal
I}^{\flat,pf}_{\triangle}(P_1;p_2;p_3) &\simeq&{2\ell\cdot
P_1-P_1^2\over\ell^2(-2\ell\cdot P_1+P_1^2)P_1^2(2\ell\cdot
p_3)}-{2\ell\cdot p_{23}-p_{23}^2\over \ell^2(-2\ell\cdot
p_{23}+p_{23}^2) p_{23}^2(2\ell\cdot p_2)}\nn
&= & {-1\over\ell^2P_1^2(2\ell\cdot p_3)}-{-1\over \ell^2
p_{23}^2(2\ell\cdot p_2)}\simeq 0~,~~~\eea
where the last line is a scale-free expression which integrates to
zero. This is the first non-trivial example showing that, although
the ${\cal Q}$-cut representation of a loop integrand is different
from the one given by Feynman diagrams, their integrated results
still match. Hence upon integration, we have the following two
equivalent expressions
\bea {1\over \ell^2(\ell-P_1)^2(\ell+p_3)^2}\simeq {\ell\cdot
P_1\over\ell^2(-2\ell\cdot P_1+P_1^2)P_1^2(\ell\cdot
p_3)}-{\ell\cdot p_{23}\over \ell^2(-2\ell\cdot p_{23}+p_{23}^2)
p_{23}^2(\ell\cdot p_2)}~,~~~\nonumber\eea
which can be used in later comparison.

~\\
{\bf Scalar integrand of box topology:} Let us focus on the
expression
\bea {\cal I}^{\flat}_{\Box}(p_1;p_2;p_3;p_4)={1\over
\ell^2(\ell-p_1)^2(\ell-p_{12})^2(\ell+p_4)^2}~.~~~\label{F-box}\eea
Applying the partial fraction identity \eref{PF} and shifting the
loop momentum, we get
\bea {\cal I}_{\Box}^{\flat,pf}
&\simeq&{1\over\ell^2}\Big({1\over(-2\ell\cdot p_1)(-2\ell\cdot
p_{12}+p_{12}^2)(2\ell\cdot p_4)}+{1\over(2\ell\cdot
p_1)(-2\ell\cdot p_2)(2\ell\cdot p_{41}+p_{41}^2)}\nn &
&~~~~~~~~~~+{1\over(2\ell\cdot p_{12}+p_{12}^2)(2\ell\cdot
p_2)(-2\ell\cdot p_3)} +{1\over(-2\ell\cdot p_4)(-2\ell\cdot
p_{41}+p_{41}^2)(2\ell\cdot p_3)}\Big)~.~~~\label{boxQF}\eea

For this example, we can directly use formula (\ref{q-cut}) to write
the ${\cal Q}$-cut representation of expression \eref{F-box}, which
is given by
\bea {\cal I}^{\flat,{\cal Q}}_{\Box}(p_1;p_2;p_3;p_4)&\simeq
&{\ell\cdot p_{12}\over(\ell\cdot
p_1)p_{12}^2}{1\over\ell^2(-2\ell\cdot p_{12}+p_{12}^2)}{\ell\cdot
p_{34}\over(\ell\cdot
p_4)p_{34}^2}+\cyclic{p_1,p_2,p_3,p_4}~.~~~\label{boxQ} \eea
Again, it is easy to figure out that the difference of \eref{boxQ}
and \eref{boxQF} is a scale-free expression\footnote{It is easy to
find the difference of the first terms of \eref{boxQ} and
\eref{boxQF}, which is $ {-(p_{12}^2+2\ell\cdot
p_{12})\over\ell^2(2\ell\cdot p_1)p_{12}^2(2\ell\cdot
p_4)p_{34}^2}$.}. So although the expression ${\cal I}^{\flat,{\cal
Q}}_{\Box}(p_1;p_2;p_3;p_4)$ is apparently different from ${\cal
I}^{\flat,pf}_{\Box}(p_1;p_2;p_3;p_4)$, they are equivalent upon
integration. Hence we have the following two equivalent expressions
\bea {1\over \ell^2(\ell-p_1)^2(\ell-p_{12})^2(\ell+p_4)^2}\simeq
{\ell\cdot p_{12}\over(\ell\cdot
p_1)p_{12}^2}{1\over\ell^2(-2\ell\cdot p_{12}+p_{12}^2)}{\ell\cdot
p_{34}\over(\ell\cdot
p_4)p_{34}^2}+\cyclic{p_1,p_2,p_3,p_4}~,~~~\nonumber\eea
which can be used in later comparison.

%%%%%%%%%%%%%%%%%
\subsection{Color-ordered 4-point amplitude in $\phi^4$ theory}
\label{subsecScalar1}
%%%%%%%%%%%%%%%%

Now, let us consider the one-loop amplitudes in scalar field theory,
and the first one is the color-ordered $\phi^4$ theory. The
analysis will be presented as follows. Firstly we compute the loop
integrand by Feynman diagrams, next we compute the loop integrand by
$\mathcal{Q}$-cut construction. Then we will compare these two integrands
directly and show their equivalence upon loop integration,
followed by a discussion in the context of unitarity cut, which
provides another independent cross-check for its validity.

By color-ordered Feynman rules, the integrand of the
4-point one-loop amplitude gets contribution from two bubble
diagrams, which is
\bea {\cal I}^{\cal F}(\ell)={1\over \ell^2(\ell-p_{12})^2}+{1\over
\ell^2(\ell-p_{41})^2}~,~~~\label{colorphi4F1} \eea
while using the ${\cal Q}$-cut construction \eref{q-cut}, it is given by

\bea {\cal I}^{\cal Q}&=&{\cal A}_L(1,2,\WH\ell_R,-\WH\ell_L){1\over
\ell^2(-2\ell\cdot p_{12}+p_{12}^2)}{\cal
A}_R(\WH\ell_L,-\WH\ell_R,3,4)+\cyclic{p_1,p_2,p_3,p_4}~.~~~\eea
where $\WH\ell=\alpha(\ell+\eta)$ ($\eta$ is determined by
$\ell^2-\eta^2=0$), and $\alpha$ is the pole's location specific to
that cut. Nonetheless, for $\phi^4$ theory the 4-point
tree amplitude is trivially $1$, i.e., ${\cal A}_L={\cal
A}_R=1$. Thus we get
\bea {\cal I}^{\cal Q}(\ell)={1\over \ell^2(-2\ell\cdot
p_{12}+p_{12}^2)}+{1\over \ell^2(-2\ell\cdot
p_{23}+p_{23}^2)}+{1\over \ell^2(-2\ell\cdot
p_{34}+p_{34}^2)}+{1\over \ell^2(-2\ell\cdot
p_{41}+p_{41}^2)}~.~~~\label{colorphi4Q} \eea
Obtaining \eref{colorphi4F1} and \eref{colorphi4Q}, we can
directly compare them. Note that each term in
\eref{colorphi4F1} is a standard bubble integrand already known
in \eref{F-bubble}, using result \eref{bubbleQ} it is easy to see
that the first term in \eref{colorphi4F1} is equivalent to the sum
of the first and third terms in \eref{colorphi4Q}, while the second
term in \eref{colorphi4F1} is equivalent to the sum of the second
and fourth terms in \eref{colorphi4Q}. Hence the one-to-one
correspondence obviously ensures the equivalence ${\cal I}^{\cal F}=
{\cal I}^{\cal Q}$.

Now we consider the unitarity cut for $s_{12}$-channel. The
traditional unitarity cut method gives
\bea \Delta^{\mathcal{F}}_{s_{12}}=\int d^4\ell_L d^4\ell_R
~\delta^4(\ell_R-\ell_L+p_{12})\delta^{+}(\ell_L^2)\delta^{+}(\ell_R^2)\times
\cA_L(1,2,\ell_R,-\ell_L)\cA_R(\ell_L,-\ell_R,3,4)~,~~~\label{unitarityF}\eea
while the contribution from the loop integrand of ${\cal Q}$-cut representation
is given by
\bea \Delta^{\mathcal{Q}}_{s_{12}}=\int d^4\ell_L d^4\ell_R
~\delta^4(\ell_R-\ell_L+p_{12})\delta^{+}(\ell_L^2)\delta^{+}(\ell_R^2)\times
\cI^{\cal Q}(\ell)~\ell_L^2\ell_R^2~.~~~\label{unitarityQ}\eea
Let us evaluate \eref{unitarityF} first. Since the 4-point tree
amplitudes are simply $1$, we have $\cA_L\cA_R=1$. When we integrate
over $\ell_R$ along with the momentum conservation delta function,
$\Delta^{\mathcal{F}}_{s_{12}}$ becomes
\bea \Delta^{\mathcal{F}}_{s_{12}}[\ell_L]=\int d^4\ell_L~
\delta^{+}(\ell_L^2)\delta^{+}((\ell_L-p_{12})^2)~,~~~\eea
on the other hand, when we integrate over $\ell_L$, $\Delta^{\mathcal{F}}_{s_{12}}$
becomes
\bea \Delta^{\mathcal{F}}_{s_{12}}[\ell_R]=\int d^4\ell_R~
\delta^{+}(\ell_R^2)\delta^{+}((\ell_R+p_{12})^2)~.~~~\eea
Now we evaluate \eref{unitarityQ}. Identifying $\ell\to \ell_L$
and integrate over $\ell_R$ against momentum conservation, we get
\bea \Delta^{\mathcal{Q}}_{s_{12}}[\ell_L]=\int d^4\ell_L
~\delta^{+}(\ell_L^2)\delta^{+}((\ell_L-p_{12})^2)\times
\ell_L^2(\ell_L-p_{12})^2~\cI^{\mathcal{Q}}(\ell_L)~.~~~\eea
Due to the remaining two delta functions, we have $\ell_L^2=0$
as well as $(\ell_L-p_{12})^2=0\to -2\ell_L\cdot p_{12}+p_{12}^2=0$.
Thus among the four terms of $\cI^{\mathcal{Q}}$ in
(\ref{colorphi4Q}) when multiplied by $\ell_L^2(\ell_L-p_{12})^2$,
only the first term survives (which equals to 1) and the rest three
terms vanish. Hence we have
\bea \Delta^{\mathcal{Q}}_{s_{12}}[\ell_L]=\int d^4\ell_L
~\delta^{+}(\ell_L^2)\delta^{+}((\ell_L-p_{12})^2)~.~~~\eea
Similarly, identifying $\ell\to \ell_R$ and integrating over $\ell_L$
against momentum conservation, we have
\bea \Delta^{\mathcal{Q}}_{s_{12}}[\ell_R]&= & \int d^4\ell_R
~\delta^{+}(\ell_R^2)\delta^{+}((\ell_R+p_{12})^2)\times
\ell_R^2(\ell_R+p_{12})^2~\cI^{\mathcal{Q}}(\ell_R)\nn
&= & \int d^4\ell_R
~\delta^{+}(\ell_R^2)\delta^{+}((\ell_R+p_{12})^2)~.~~~\eea
where only the third term in (\ref{colorphi4Q}) survives, when
multiplied by $\ell_R^2(\ell_R+p_{12})^2$ upon on-shell
conditions $\ell_R^2=0$, $(\ell_R+p_{12})^2=0$. The explicit
calculation above verifies the equivalence
$\Delta^{\mathcal{Q}}_{s_{12}}[\ell_L]=\Delta^{\mathcal{F}}_{s_{12}}[\ell_L]$
and
$\Delta^{\mathcal{Q}}_{s_{12}}[\ell_R]=\Delta^{\mathcal{F}}_{s_{12}}[\ell_R]$
in the $s_{12}$-channel unitarity cut. This example also
clearly demonstrates how the two terms in ${\cal Q}$-cut
representation correspond to one standard unitarity cut.

The $s_{14}$-channel unitarity cut has no essential difference
with the $s_{12}$-channel, and the equivalence between two loop integrands
generated by Feynman diagrams and
$\mathcal{Q}$-cut construction is valid for all cut channels.

%%%%%%%%%%%%%%%%%
\subsection{Color-ordered 4-point amplitude in $\phi^3$ theory}
\label{subsecScalar2}
%%%%%%%%%%%%%%%%

%
\begin{figure}
\centering
  \includegraphics[width=6.5in]{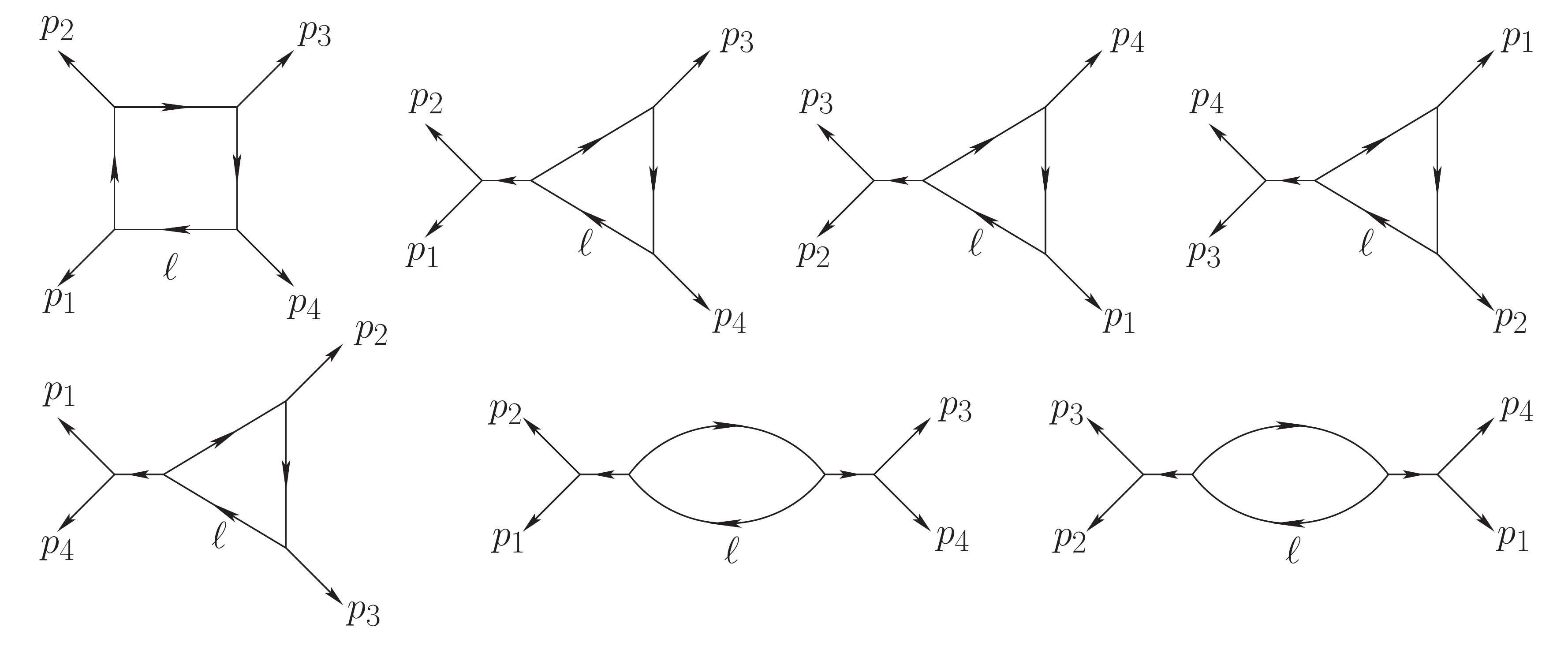}\\
  \caption{The contributing Feynman diagrams of color-ordered 4-point one-loop amplitude
  in $\phi^3$ theory.}\label{CO4PT}
\end{figure}

Now we turn to a more complicated example, namely the 4-point one-loop
amplitude in $\phi^3$ theory.

Directly from Feynman diagrams, the loop integrand is
\bea {\cal I}^{\cal F}&=&{\cal I}_{\Box}^{\phi^3}+{\cal
I}_{\triangle}^{\phi^3}+{\cal I}_{\between}^{\phi^3}~,~~~\label{4F1}
\eea
where
\bea &&{\cal I}_{\Box}^{\phi^3}={1\over
\ell^2(\ell-p_1)^2(\ell-p_{12})^2(\ell+p_4)^2}~~,~~{\cal
I}_{\triangle}^{\phi^3}={1\over
\ell^2(\ell-p_{12})^2(\ell+p_4)^2p_{12}^2}+\cyclic{p_1,p_2,p_3,p_4}~,\nn
&&{\cal I}_{\between}^{\phi^3}={1\over
\ell^2(\ell-p_{12})^2(p_{12}^2)^2}+{1\over
\ell^2(\ell-p_{41})^2(p_{41}^2)^2}~.~~~\eea
Given by ${\cal Q}$-cut construction, the loop integrand is
\bea {\cal I}^{\mathcal{Q}}&=&{\cal
A}_L(1,2,\WH\ell_R,-\WH\ell_L){1\over \ell^2(-2\ell\cdot
p_{12}+p_{12}^2)}{\cal
A}_R(\WH\ell_L,-\WH\ell_R,3,4)+\cyclic{p_1,p_2,p_3,p_4}~.~~~\eea
where
\bea {\cal A}_L(i,j,\WH\ell_R,-\WH\ell_L)={1\over
(\WH\ell_L-p_i)^2}+{1\over p_{ij}^2}={-\ell_L\cdot p_{ij}\over
p_{ij}^2(\ell_L\cdot p_i)}+{1\over p_{ij}^2}~.~~~\eea
Thus we have
\bea {\cal I}^{\mathcal{Q}}[\ell] &=&\Big({-\ell\cdot p_{12}\over
p_{12}^2(\ell\cdot p_1)}+{1\over p_{12}^2}\Big){1\over
\ell^2(-2\ell\cdot p_{12}+p_{12}^2)}\Big({-\ell\cdot p_{34}\over
p_{34}^2(\ell\cdot p_4)}+{1\over
p_{34}^2}\Big)+\cyclic{p_1,p_2,p_3,p_4}~.~~~\label{4Q} \eea
By expanding this, we get various terms with different numbers of linear
propagators. Those with only one linear propagator $(-2\ell\cdot
p_{12}+p_{12}^2)$ in the denominator must come from bubble diagrams,
while those with the additional denominator $(\ell\cdot p_i)$ come
from triangle diagrams. Finally, terms with three linear
propagators must come from box diagrams.

Let us continue to directly compare the two results above. With
the warmup exercises, this comparison is
straightforward. The one-to-one correspondence at the
integrand-level can be seen by applying \eref{boxQ},
\eref{triangleQ} and \eref{bubbleQ}, to \eref{4F1} which is from
Feynman diagrams. For the box diagrams, with \eref{boxQ} we
have
\bea {\cal I}_{\Box}^{\phi^3}\simeq {\ell\cdot p_{12}\over(\ell\cdot
p_1)p_{12}^2}{1\over\ell^2(-2\ell\cdot p_{12}+p_{12}^2)}{\ell\cdot
p_{34}\over(\ell\cdot
p_4)p_{34}^2}+\cyclic{p_1,p_2,p_3,p_4}~.~~~\eea
The appearance of these terms in (\ref{4Q}) is clear. Next for
the triangle diagrams, with \eref{triangleQ} we have
\bea {\cal I}_{\triangle}^{\phi^3}&\simeq& {\ell\cdot
p_{12}\over\ell^2(-2\ell\cdot p_{12}+p_{12}^2)(\ell\cdot
p_4)(p_{12}^2)^2}-{\ell\cdot p_{34}\over \ell^2(-2\ell\cdot
p_{34}+p_{34}^2) (\ell\cdot
p_3)(p_{34}^2)^2}+\cyclic{p_1,p_2,p_3,p_4}\nonumber\\
&=&{\ell\cdot p_{12}\over\ell^2(-2\ell\cdot
p_{12}+p_{12}^2)(\ell\cdot p_4)(p_{12}^2)^2}-{\ell\cdot p_{12}\over
\ell^2(-2\ell\cdot p_{12}+p_{12}^2) (\ell\cdot
p_1)(p_{12}^2)^2}+\cyclic{p_1,p_2,p_3,p_4}~,~~~\eea
where in the second line we have replaced
$\{p_1,p_2,p_3,p_4\}\to \{p_3,p_4,p_1,p_2\}$ for the second term,
which causes no difference due to the cyclic invariance of
$\cI_{\triangle}^{\phi^3}$. Then the one-to-one correspondence of
these eight terms to those in (\ref{4Q}) is also clear.
Finally for the bubble diagrams, with \eref{bubbleQ} we have
\bea {\cal I}^{\phi^3}_{\between}&\simeq&{1\over \ell^2(-2\ell\cdot
p_{12}+p_{12}^2)(p_{12}^2)^2}+{1\over \ell^2(-2\ell\cdot
p_{34}+p_{34}^2)(p_{34}^2)^2}\nonumber\\
&&~~~~~~~~+{1\over \ell^2(-2\ell\cdot
p_{41}+p_{41}^2)(p_{41}^2)^2}+{1\over \ell^2(-2\ell\cdot
p_{23}+p_{23}^2)(p_{23}^2)^2}~,~~~\label{phi3bubQ} \eea
which match the remaining four terms in (\ref{4Q}). Hence we have shown
the equivalence ${\cal I}^{\mathcal{F}}\simeq {\cal
I}^{\mathcal{Q}}$ term-by-term.

After the direct comparison at the integrand-level, we now move to
the integral-level under the traditional unitarity cut. Again we take
$s_{12}$-channel as an example, and the unitarity cut
yields
\bea \Delta^{\mathcal{F}}_{s_{12}}&=&\int d^4\ell_L d^4\ell_R
~\delta^4(\ell_R-\ell_L+p_{12})\delta^{+}(\ell_L^2)\delta^{+}(\ell_R^2)\times
\cA_L(1,2,\ell_R,-\ell_L)\cA_R(\ell_L,-\ell_R,3,4)\nonumber\\
&=&\int d^4\ell_L d^4\ell_R
~\delta^4(\ell_R-\ell_L+p_{12})\delta^{+}(\ell_L^2)\delta^{+}(\ell_R^2)\Big({1\over
(\ell_L-p_1)^2}+{1\over p_{12}^2}\Big)\Big({1\over
(\ell_R-p_3)^2}+{1\over p_{34}^2}\Big)~.~~~\eea
Integrating over $\ell_R$ first, we get
\bea \Delta^{\mathcal{F}}_{s_{12}}[\ell_L]&=&\int
d^4\ell_L~\delta^{+}(\ell_L^2)\delta^{+}((\ell_L-p_{12})^2)\Big({1\over
(\ell_L-p_1)^2}+{1\over p_{12}^2}\Big)\Big({1\over
(\ell_L+p_4)^2}+{1\over p_{34}^2}\Big)~,~~~\eea
while integrating over $\ell_L$ first, we get
\bea \Delta^{\mathcal{F}}_{s_{12}}[\ell_R]&=&\int d^4\ell_R
~\delta^{+}(\ell_R^2)\delta^{+}((\ell_R+p_{12})^2)\Big({1\over
(\ell_R+p_2)^2}+{1\over p_{12}^2}\Big)\Big({1\over
(\ell_R-p_3)^2}+{1\over p_{34}^2}\Big)~.~~~\eea
Now we repeat the calculation with the ${\cal Q}$-cut representation \eref{4Q}.
Identifying $\ell=\ell_L$ and integrating over $\ell_R$ against momentum conservation, we
get
\bea \Delta^{\mathcal{Q}}_{s_{12}}[\ell_L]&=&\int d^4\ell_L
~\delta^{+}(\ell_L^2)\delta^{+}((\ell_L-p_{12})^2)\times
\ell_L^2(-2\ell_L\cdot
p_{12}+p_{12}^2)~\cI_Q(\ell_L)\nonumber\\
&=&\int d^4\ell_L
~\delta^{+}(\ell_L^2)\delta^{+}((\ell_L-p_{12})^2)\Big({-\ell_L\cdot
p_{12}\over p_{12}^2(\ell_L\cdot p_1)}+{1\over
p_{12}^2}\Big)\Big({-\ell_L\cdot p_{34}\over p_{34}^2(\ell_L\cdot
p_4)}+{1\over p_{34}^2}\Big)\nonumber\\
&=&\int d^4\ell_L
~\delta^{+}(\ell_L^2)\delta^{+}((\ell_L-p_{12})^2)\Big(-{1\over
2\ell_L\cdot p_1}+{1\over p_{12}^2}\Big)\Big({1\over 2\ell_L\cdot
p_4}+{1\over p_{34}^2}\Big)~,~~~\eea
where in the last line we have used the on-shell conditions
$2\ell_L\cdot p_{12}=p_{12}^2$. Similarly, identifying $\ell=\ell_R$ and
integrating over $\ell_L$ against momentum conservation, we get
\bea \Delta^{\mathcal{Q}}_{s_{12}}[\ell_R]&=&\int d^4\ell_R
~\delta^{+}(\ell_R^2)\delta^{+}((\ell_R+p_{12})^2)\times
\ell_R^2(-2\ell_R\cdot
p_{34}+p_{34}^2)~\cI_Q(\ell_R)\nonumber\\
&=&\int d^4\ell_R
~\delta^{+}(\ell_R^2)\delta^{+}((\ell_R+p_{12})^2)\Big({-\ell_R\cdot
p_{34}\over p_{34}^2(\ell_R\cdot p_3)}+{1\over
p_{34}^2}\Big)\Big({-\ell_R\cdot p_{12}\over p_{12}^2(\ell_R\cdot
p_2)}+{1\over p_{12}^2}\Big)\nonumber\\
&=&\int d^4\ell_R
~\delta^{+}(\ell_R^2)\delta^{+}((\ell_R+p_{12})^2)\Big(-{1\over
2\ell_R\cdot p_3}+{1\over p_{34}^2}\Big)\Big({1\over 2\ell_R\cdot
p_2}+{1\over p_{12}^2}\Big)~,~~~\eea
where in the last line we have used the on-shell conditions
$2\ell_R\cdot p_{12}=-p_{12}^2$. The cut condition
$\ell_L^2=0$ or $\ell_R^2=0$ and the massless condition $p_i^2=0$
immediately imply $\Delta^{\mathcal{Q}}_{s_{12}}[\ell_L]\simeq
\Delta^{\mathcal{F}}_{s_{12}}[\ell_L]$ and
$\Delta^{\mathcal{Q}}_{s_{12}}[\ell_R]\simeq
\Delta^{\mathcal{F}}_{s_{12}}[\ell_R]$.

Similar calculation can be done for $s_{14}$-channel by cyclic
shift $p_i\to p_{i+1}$. Again one can show that the result
will be the same no matter which integrand, namely $\cI^{\cal F}$ or
$\cI^{\cal Q}$, is used.

%%%%%%%%%%%%%%%%%
\subsection{Non-color-ordered 4-point amplitude in $\phi^4$ theory}
\label{subsecScalar3}
%%%%%%%%%%%%%%%%

Now, we discuss the non-color-ordered amplitudes. Different from the
color-ordered ones, in principle all possible physical poles in terms of
Lorentz invariants could appear in the integrand.
Furthermore, special attention should be paid to the symmetry
factor, in order to avoid over-counting terms. In this subsection, we
consider scalars with the standard interaction term
$\phi^4/4!$.

The loop integrand gets contributions from three bubble diagrams, and
by Feynman rules it can be written as
\bea \cI^{\mathcal{F}}={1\over
2}\cI_{\between}^{\flat}(p_1,p_2;p_3,p_4)+{1\over
2}\cI_{\between}^{\flat}(p_1,p_3;p_2,p_4)+{1\over
2}\cI_{\between}^{\flat}(p_1,p_4;p_2,p_3)~,~~~\eea
where the semicolon separates two pairs of external momenta onto two ends of
the bubble diagram, and $\cI_{\between}^{\flat}(p_1,p_2;p_3,p_4) ={1\over
\ell^2 (\ell-p_{12})^2}$. The prefactor $1/2$ is the non-color-ordered
symmetry factor.

Then we shall write down the loop integrand by the ${\cal Q}$-cut
construction \eref{q-cut}. Since there is no color structure, we
need to sum over all possible $P_L$ cuts where $P_L=p_{12}$,
$p_{13}$, $p_{14}$, $p_{23}$, $p_{24}$, $p_{34}$. Thus there will be
six $\mathcal{Q}$-cut terms, compared to four of the color-ordered
case. Furthermore, a symmetry factor $1/2$ should be associated to
each $\mathcal{Q}$-cut term, since in the construction we have
considered all possible orderings of external legs when calculating
tree amplitudes. When sewing the two legs denoted by
$\widehat{\ell}_L, \widehat{\ell}_R$ of two tree amplitudes, we take all
four different combinations into account, and clearly two of them
represent exactly the same diagrams as the rest two. So when we use
the non-color-ordered tree amplitudes as inputs, the symmetry factor
$1/2$ should be introduced to offset this over-counting.

Having considered the subtlety above, the loop integrand of ${\cal Q}$-cut
representation is given by
\bea {\cal I}^{\mathcal{Q}}&=&{1\over2}{\cal
A}_L(1,2,\WH\ell_R,-\WH\ell_L){1\over \ell^2(-2\ell\cdot
p_{12}+p_{12}^2)}{\cal A}_R(\WH\ell_L,-\WH\ell_R,3,4)\nn
& &+{1\over2}{\cal A}_L(1,3,\WH\ell_R,-\WH\ell_L){1\over
\ell^2(-2\ell\cdot p_{13}+p_{13}^2)}{\cal
A}_R(\WH\ell_L,-\WH\ell_R,2,4)\nn
& &+{1\over2}{\cal A}_L(1,4,\WH\ell_R,-\WH\ell_L){1\over
\ell^2(-2\ell\cdot p_{14}+p_{14}^2)}{\cal
A}_R(\WH\ell_L,-\WH\ell_R,2,3)\nn
& &+{1\over2}{\cal A}_L(2,3,\WH\ell_R,-\WH\ell_L){1\over
\ell^2(-2\ell\cdot p_{23}+p_{23}^2)}{\cal
A}_R(\WH\ell_L,-\WH\ell_R,1,4)\nn
& &+{1\over2}{\cal A}_L(2,4,\WH\ell_R,-\WH\ell_L){1\over
\ell^2(-2\ell\cdot p_{24}+p_{24}^2)}{\cal
A}_R(\WH\ell_L,-\WH\ell_R,1,3)\nn
 & &+{1\over2}{\cal
A}_L(3,4,\WH\ell_R,-\WH\ell_L){1\over \ell^2(-2\ell\cdot
p_{34}+p_{34}^2)}{\cal A}_R(\WH\ell_L,-\WH\ell_R,1,2)~,~~~\eea
where the particle labels in the tree amplitudes just indicate
external momenta, without color ordering. Since ${\cal A}_L={\cal
A}_R=1$ for all channels, we get
\bea {\cal I}^{\mathcal{Q}} &=&{1\over 2\ell^2(-2\ell\cdot
p_{12}+p_{12}^2)}+{1\over 2\ell^2(-2\ell\cdot
p_{13}+p_{13}^2)}+{1\over 2\ell^2(-2\ell\cdot p_{14}+p_{14}^2)}\nn
& &+{1\over 2\ell^2(-2\ell\cdot
p_{23}+p_{23}^2)}+{1\over 2\ell^2(-2\ell\cdot
p_{24}+p_{24}^2)}+{1\over 2\ell^2(-2\ell\cdot
p_{34}+p_{34}^2)}~.~~~\eea
Next, let us directly compare two integrands with
\eref{bubbleQ}. It is easy to work out that
\bea &&{1\over 2}\cI_{\between}^{\flat}(p_1,p_2;p_3,p_4) \simeq
{1\over 2\ell^2(-2\ell\cdot p_{12}+p_{12}^2)}+{1\over
2\ell^2(-2\ell\cdot p_{34}+p_{34}^2)}~,~~~\nn
&&{1\over 2}\cI_{\between}^{\flat}(p_1,p_3;p_2,p_4) \simeq {1\over
2\ell^2(-2\ell\cdot p_{13}+p_{13}^2)}+{1\over 2\ell^2(-2\ell\cdot
p_{24}+p_{24}^2)}~,~~~\nn
&&{1\over 2}\cI_{\between}^{\flat}(p_1,p_4;p_2,p_3) \simeq {1\over
2\ell^2(-2\ell\cdot p_{14}+p_{14}^2)}+{1\over 2\ell^2(-2\ell\cdot
p_{23}+p_{23}^2)}~.~~~\eea
Therefore the equivalence ${\cal I}^{\mathcal{F}}\simeq {\cal
I}^{\mathcal{Q}}$ is clear.

Then we compare these two integrands at the integral-level by using the
unitarity cut. Again, let us take $s_{12}$-channel as an
example. Following the definitions (\ref{unitarityF}) and
(\ref{unitarityQ}), the computation is exactly the same as that for
color-ordered amplitudes of $\phi^4$ theory (it is not
surprising since in both cases the 4-point tree amplitudes are
1). So we have
\bea
&&\Delta^{\mathcal{Q}}_{s_{12}}[\ell_L]\simeq\Delta^{\mathcal{F}}_{s_{12}}[\ell_L]={1\over
2}\int
d^4\ell_L~ \delta^{+}(\ell_L^2)\delta^{+}((\ell_L-p_{12})^2)~,~~~\nonumber\\
&&\Delta^{\mathcal{Q}}_{s_{12}}[\ell_R]\simeq\Delta^{\mathcal{F}}_{s_{12}}[\ell_R]={1\over
2}\int d^4\ell_R~
\delta^{+}(\ell_R^2)\delta^{+}((\ell_R+p_{12})^2)~.~~~\nonumber\eea
The same analysis for $s_{13}$- and $s_{14}$-cut channels shows
that
$\Delta^{\mathcal{Q}}_{s_{13}}[\ell_L]\simeq\Delta^{\mathcal{F}}_{s_{13}}[\ell_L]$,
$\Delta^{\mathcal{Q}}_{s_{13}}[\ell_R]\simeq\Delta^{\mathcal{F}}_{s_{13}}[\ell_R]$
as well as
$\Delta^{\mathcal{Q}}_{s_{14}}[\ell_L]\simeq\Delta^{\mathcal{F}}_{s_{14}}[\ell_L]$,
$\Delta^{\mathcal{Q}}_{s_{14}}[\ell_R]\simeq\Delta^{\mathcal{F}}_{s_{14}}[\ell_R]$.
Hence the equivalence $\cI^{\mathcal{Q}}\simeq \cI^{\mathcal{F}}$
holds for each unitarity cut channel.

%%%%%%%%%%%%%%%%%
\subsection{Non-color-ordered 4-point amplitude in $\phi^3$ theory}
\label{subsecScalar4}
%%%%%%%%%%%%%%%%

%
\begin{figure}
\centering
  \includegraphics[width=6.5in]{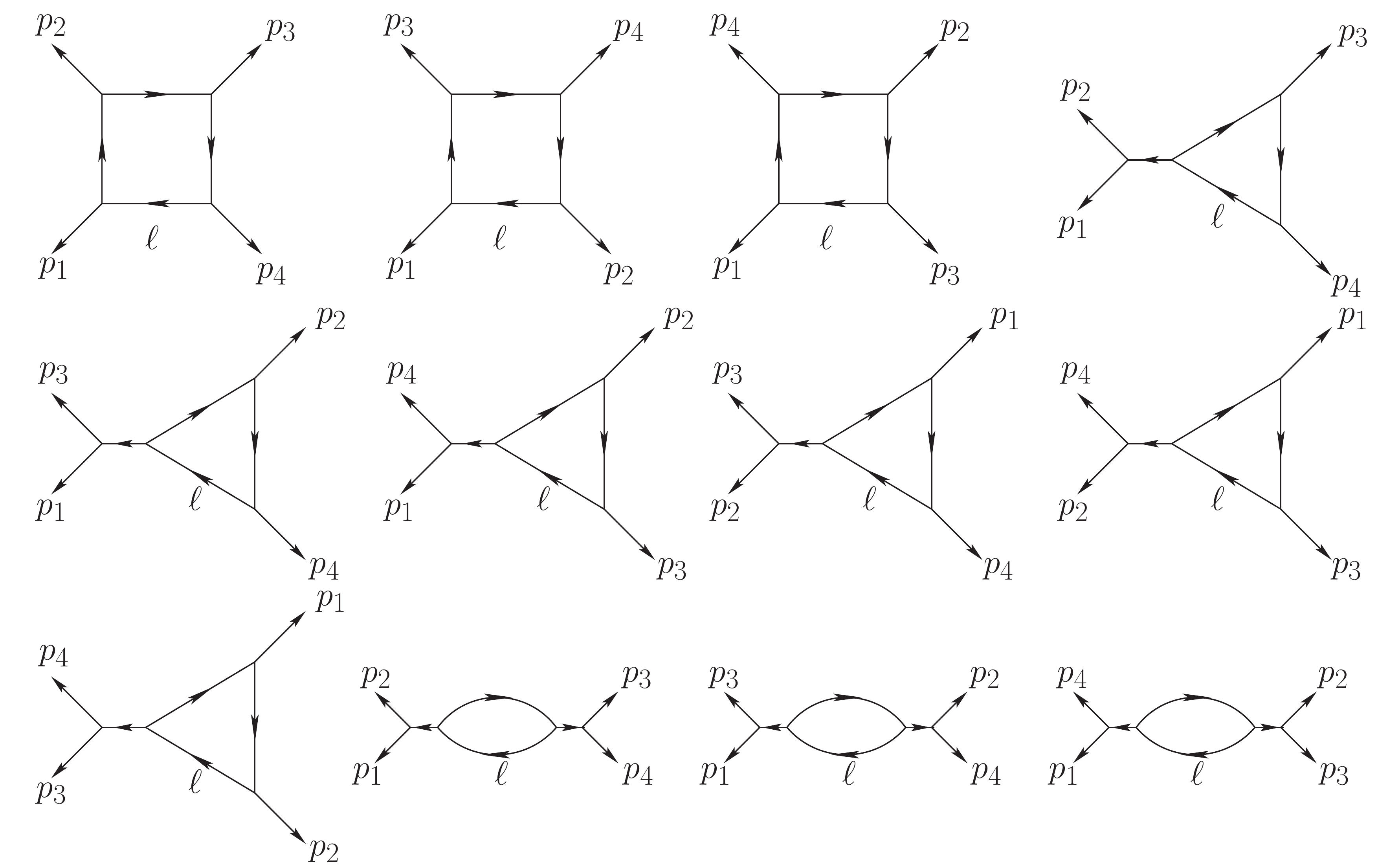}\\
  \caption{The contributing Feynman diagrams of non-color-ordered 4-point one-loop
  amplitude in $\phi^3$ theory.}\label{NCO4PT}
\end{figure}

The non-color-ordered 4-point one-loop amplitude in
$\phi^3$ theory is more complicated, yet the discussion follows the
same way as in previous subsections. The
interaction term here is ${\phi^3/ 3!}$. The loop integrand gets
contributions from three box diagrams, six triangle diagrams and
three bubble diagrams, as shown in Figure (\ref{NCO4PT}), given by
\bea
\cI^{\mathcal{F}}=\cI^{\phi^3}_{\Box}+\cI^{\phi^3}_{\triangle}+\cI^{\phi^3}_{\between}~,~~~\label{nonCphi3F}\eea
where
\bea &&{\cal I}_{\Box}^{\phi^3}=\cI^{\flat}_{\Box}(p_1;p_2;p_3;p_4)+\cI^{\flat}_{\Box}(p_1;p_3;p_4;p_2)
+\cI^{\flat}_{\Box}(p_1;p_4;p_2;p_3)~,~~~\\
&&{\cal
I}_{\triangle}^{\phi^3}=\cI^{\flat}_{\triangle}(p_1,p_2;p_3;p_4)+\cI^{\flat}_{\triangle}
(p_1,p_3;p_2;p_4)+\cI^{\flat}_{\triangle}(p_1,p_4;p_2;p_3)\nonumber\\
&&~~~~~~~+\cI^{\flat}_{\triangle}(p_2,p_3;p_1;p_4)+\cI^{\flat}_{\triangle}(p_2,p_4;p_1;p_3)
+\cI^{\flat}_{\triangle}(p_3,p_4;p_1;p_2)~,~~~\\
&&{\cal I}_{\between}^{\phi^3}={1\over
2}\cI^{\flat}_{\between}(p_1,p_2;p_3,p_4)+{1\over
2}\cI^{\flat}_{\between}(p_1,p_3;p_2,p_4)+{1\over
2}\cI^{\flat}_{\between}(p_1,p_4;p_2,p_3)~.~~~\eea
and the expressions of $\cI^{\flat}_{\Box}(p_1;p_2;p_3;p_4)$,
$\cI^{\flat}_{\triangle}(p_1,p_2;p_3;p_4)$,
$\cI^{\flat}_{\between}(p_1,p_2;p_3,p_4)$ can be inferred from
\eref{F-box}, \eref{F-tri} and  \eref{F-bubble}. Again the semicolon
separates external momenta consecutively onto each end of
corresponding Feynman diagrams, and a symmetry factor $1/2$ is
associated to each bubble diagram. Note that for box diagrams (or
triangles), a diagram $\cI_{\Box}^{\flat}(p_1;p_2;p_3;p_4)$
is topologically equivalent to its mirror reflection
$\cI_{\Box}^{\flat}(p_1;p_4;p_3;p_2)$ (or
$\cI_{\triangle}^{\flat}(p_1;p_2;p_3)$ to
$\cI_{\triangle}^{\flat}(p_1;p_3;p_2)$), so we must not
over-count them.

The integrand in $\mathcal{Q}$-cut representation is given by
\bea {\cal I}^{\mathcal{Q}}&=&{1\over2}{\cal
A}_L(1,2,\WH\ell_R,-\WH\ell_L){1\over \ell^2(-2\ell\cdot
p_{12}+p_{12}^2)}{\cal A}_R(\WH\ell_L,-\WH\ell_R,3,4)\nn
& &+{1\over2}{\cal A}_L(1,3,\WH\ell_R,-\WH\ell_L){1\over
\ell^2(-2\ell\cdot p_{13}+p_{13}^2)}{\cal
A}_R(\WH\ell_L,-\WH\ell_R,2,4)\nn
& &+{1\over2}{\cal A}_L(1,4,\WH\ell_R,-\WH\ell_L){1\over
\ell^2(-2\ell\cdot p_{14}+p_{14}^2)}{\cal
A}_R(\WH\ell_L,-\WH\ell_R,2,3)\nn
& &+{1\over2}{\cal A}_L(2,3,\WH\ell_R,-\WH\ell_L){1\over
\ell^2(-2\ell\cdot p_{23}+p_{23}^2)}{\cal
A}_R(\WH\ell_L,-\WH\ell_R,1,4)\nn
& &+{1\over2}{\cal A}_L(2,4,\WH\ell_R,-\WH\ell_L){1\over
\ell^2(-2\ell\cdot p_{24}+p_{24}^2)}{\cal
A}_R(\WH\ell_L,-\WH\ell_R,1,3)\nn
& &+{1\over2}{\cal A}_L(3,4,\WH\ell_R,-\WH\ell_L){1\over
\ell^2(-2\ell\cdot p_{34}+p_{34}^2)}{\cal
A}_R(\WH\ell_L,-\WH\ell_R,1,2)~,~~~\eea
where the $D$-dimensional non-color-ordered tree amplitude is given by
\bea {\cal A}_L(i,j,\WH\ell_R,-\WH\ell_L)={-\ell_L\cdot p_{ij}\over
p_{ij}^2(\ell_L\cdot p_i)}+{-\ell_L\cdot p_{ij}\over
p_{ij}^2(\ell_L\cdot p_j)}+{1\over p_{ij}^2}~.~~~\eea
Hence we get
\bea {\cal I}^{\mathcal{Q}}&=&{1\over2}\Big({-\ell\cdot p_{12}\over
p_{12}^2(\ell\cdot p_1)}+{-\ell\cdot p_{12}\over p_{12}^2(\ell\cdot
p_2)}+{1\over p_{12}^2}\Big){1\over \ell^2(-2\ell\cdot
p_{12}+p_{12}^2)}\Big({-\ell\cdot p_{34}\over p_{34}^2(\ell\cdot
p_3)}+{-\ell\cdot p_{34}\over p_{34}^2(\ell\cdot p_4)}+{1\over
p_{34}^2}\Big)\nn
& &+{1\over2}\Big({-\ell\cdot p_{13}\over p_{13}^2(\ell\cdot
p_1)}+{-\ell\cdot p_{13}\over p_{13}^2(\ell\cdot p_3)}+{1\over
p_{13}^2}\Big){1\over \ell^2(-2\ell\cdot
p_{13}+p_{13}^2)}\Big({-\ell\cdot p_{24}\over p_{24}^2(\ell\cdot
p_2)}+{-\ell\cdot p_{24}\over p_{24}^2(\ell\cdot p_4)}+{1\over
p_{24}^2}\Big)\nn
& &+{1\over2}\Big({-\ell\cdot p_{14}\over p_{14}^2(\ell\cdot
p_1)}+{-\ell\cdot p_{14}\over p_{14}^2(\ell\cdot p_4)}+{1\over
p_{14}^2}\Big){1\over \ell^2(-2\ell\cdot
p_{14}+p_{14}^2)}\Big({-\ell\cdot p_{23}\over p_{23}^2(\ell\cdot
p_2)}+{-\ell\cdot p_{23}\over p_{23}^2(\ell\cdot p_3)}+{1\over
p_{23}^2}\Big)\nn
& &+{1\over2}\Big({-\ell\cdot p_{23}\over p_{23}^2(\ell\cdot
p_2)}+{-\ell\cdot p_{23}\over p_{23}^2(\ell\cdot p_3)}+{1\over
p_{23}^2}\Big){1\over \ell^2(-2\ell\cdot
p_{23}+p_{23}^2)}\Big({-\ell\cdot p_{14}\over p_{14}^2(\ell\cdot
p_1)}+{-\ell\cdot p_{14}\over p_{14}^2(\ell\cdot p_4)}+{1\over
p_{14}^2}\Big)\nn
& &+{1\over2}\Big({-\ell\cdot p_{24}\over p_{24}^2(\ell\cdot
p_2)}+{-\ell\cdot p_{24}\over p_{24}^2(\ell\cdot p_4)}+{1\over
p_{24}^2}\Big){1\over \ell^2(-2\ell\cdot
p_{24}+p_{24}^2)}\Big({-\ell\cdot p_{13}\over p_{13}^2(\ell\cdot
p_1)}+{-\ell\cdot p_{13}\over p_{13}^2(\ell\cdot p_3)}+{1\over
p_{13}^2}\Big)\nn
& &+{1\over2}\Big({-\ell\cdot p_{34}\over p_{34}^2(\ell\cdot
p_3)}+{-\ell\cdot p_{34}\over p_{34}^2(\ell\cdot p_4)}+{1\over
p_{34}^2}\Big){1\over \ell^2(-2\ell\cdot
p_{34}+p_{34}^2)}\Big({-\ell\cdot p_{12}\over p_{12}^2(\ell\cdot
p_1)}+{-\ell\cdot p_{12}\over p_{12}^2(\ell\cdot p_2)}+{1\over
p_{12}^2}\Big)~.~~~\label{nocolor3Q} \eea
Now let us compare these two integrands directly at the integrand-level.
As usual, we can rewrite
$\cI^{\mathcal{F}}=\cI_{\Box}^{\phi^3}+\cI_{\triangle}^{\phi^3}+\cI_{\between}^{\phi^3}$
into the $\mathcal{Q}$-cut form with \eref{boxQ},
\eref{triangleQ} and \eref{bubbleQ}. However, the one-to-one
correspondence is not manifest when in terms of the loop integrand
of the form (\ref{nonCphi3F}), since $\cI^{\mathcal{Q}}$ is
permutation invariant with respect to external momenta, while
$\cI^{\mathcal{F}}$ of the form (\ref{nonCphi3F}) is not. Recall
that a single box or triangle diagram is topologically equivalent to
its mirror reflection, we can rewrite $\cI^{\mathcal{F}}$ as
\bea &&{\cal I}_{\Box}^{\phi^3}={1\over 2}\Big(\cI^{\flat}_{\Box}(p_1;p_2;p_3;p_4)+\cI^{\flat}_{\Box}(p_1;p_3;p_4;p_2)+\cI^{\flat}_{\Box}(p_1;p_4;p_2;p_3)\nonumber\\
&&~~~~~~~~~~~~~~~~~~~~~~~~~~~~+\cI^{\flat}_{\Box}(p_1;p_4;p_3;p_2)+\cI^{\flat}_{\Box}(p_1;p_2;p_4;p_3)+\cI^{\flat}_{\Box}(p_1;p_3;p_2;p_4)\Big)~,~~~\\
&&{\cal
I}_{\triangle}^{\phi^3}={1\over 2}\Big(\cI^{\flat}_{\triangle}(p_1,p_2;p_3;p_4)+\cI^{\flat}_{\triangle}(p_1,p_3;p_2;p_4)+\cI^{\flat}_{\triangle}(p_1,p_4;p_2;p_3)\nonumber\\
&&~~~~~~~~~~~~~~~~~~~~~~~~~~~~~~~+\cI^{\flat}_{\triangle}(p_2,p_3;p_1;p_4)+\cI^{\flat}_{\triangle}(p_2,p_4;p_1;p_3)+\cI^{\flat}_{\triangle}(p_3,p_4;p_1;p_2)\nonumber\\
&&~~~~~~~~~~~~~+\cI^{\flat}_{\triangle}(p_1,p_2;p_4;p_3)+\cI^{\flat}_{\triangle}(p_1,p_3;p_4;p_2)+\cI^{\flat}_{\triangle}(p_1,p_4;p_3;p_2)\nonumber\\
&&~~~~~~~~~~~~~~~~~~~~~~~~~~~~~~~+\cI^{\flat}_{\triangle}(p_2,p_3;p_4;p_1)+\cI^{\flat}_{\triangle}(p_2,p_4;p_3;p_1)+\cI^{\flat}_{\triangle}(p_3,p_4;p_2;p_1)\Big)~,~~~\eea
while $\cI_{\between}^{\phi^3}$ remains the same. By rewriting this
one can find that, after transforming $\cI^{\mathcal{F}}$ with
\eref{boxQ}, \eref{triangleQ} and \eref{bubbleQ}, each term in
$\cI^{\mathcal{F}}$ has its correspondence in $\cI^{\mathcal{Q}}$.
An implicit evidence of the equivalence can be shown by counting the
number of terms. Expanding $\cI^{\mathcal{Q}}$ given by
(\ref{nocolor3Q}), we get 54 terms. Judged by the appearance of loop
momentum $\ell$ in the denominator of each term, we can infer that 6
terms come from bubble diagrams, $4\times 6=24$ terms come from
triangle diagrams and $4\times 6=24$ terms come from box diagrams.
Meanwhile, $\cI_{\between}^{\phi^3}$ gives $3\times 2=6$ terms,
$\cI_{\triangle}^{\phi^3}$ $12\times 2=24$ terms and
$\cI_{\Box}^{\phi^3}$ $6\times 4=24$ terms after expressed as
$\mathcal{Q}$-cut forms with \eref{boxQ}, \eref{triangleQ} and
\eref{bubbleQ}, and the symmetry factor $1/2$ is exactly expected. Of
course, the honest comparison can be done by explicitly expanding
$\cI^{\mathcal{Q}}$ and $\cI^{\mathcal{F}}$, and comparing them one by
one for all 54 terms, which indeed confirms the equivalence
$\cI^{\mathcal{F}}\simeq \cI^{\mathcal{Q}}$.

Next we turn to the comparison by the unitarity cut. Taking
$s_{12}$-cut channel as an example, for the contribution from
$\cI^{\mathcal{Q}}(\ell)$, we need to evaluate
\bea \Delta^{\mathcal{Q}}_{s_{12}}=\int d^4\ell_L d^4\ell_R
~\delta^4(\ell_R-\ell_L+p_{12})\delta^{+}(\ell_L^2)\delta^{+}(\ell_R^2)\times
\cI^{\mathcal{Q}}(\ell)~\ell_L^2\ell_R^2~.~~~\eea
Identifying $\ell=\ell_L$ (or $\ell=\ell_R$) and integrating over
$\ell_R$ (or $\ell_L$) against momentum conservation,
we get respectively {\small
\bea \Delta^{\mathcal{Q}}_{s_{12}}[\ell_L]&=&\int d^4\ell_L
~\delta^{+}(\ell_L^2)\delta^{+}((\ell_L-p_{12})^2){1\over2}\Big({-\ell_L\cdot
p_{12}\over p_{12}^2(\ell_L\cdot p_1)}+{-\ell_L\cdot p_{12}\over
p_{12}^2(\ell_L\cdot p_2)}+{1\over p_{12}^2}\Big)\Big({-\ell_L\cdot
p_{34}\over p_{34}^2(\ell_L\cdot p_3)}+{-\ell_L\cdot p_{34}\over
p_{34}^2(\ell_L\cdot p_4)}+{1\over
p_{34}^2}\Big)\nonumber\\
&=&\int d^4\ell_L
~\delta^{+}(\ell_L^2)\delta^{+}((\ell_L-p_{12})^2){1\over2}\Big(-{1\over
2\ell_L\cdot p_1}-{1\over 2\ell_L\cdot p_2}+{1\over
p_{12}^2}\Big)\Big({1\over 2\ell_L\cdot p_3}+{1\over 2\ell_L\cdot
p_4}+{1\over p_{34}^2}\Big)~,~~~\eea
}\\and {\small
\bea \Delta^{\mathcal{Q}}_{s_{12}}[\ell_R]&=&\int d^4\ell_R
~\delta^{+}(\ell_R^2)\delta^{+}((\ell_R+p_{12})^2){1\over2}\Big({-\ell_R\cdot
p_{34}\over p_{34}^2(\ell_R\cdot p_3)}+{-\ell_R\cdot p_{34}\over
p_{34}^2(\ell_R\cdot p_4)}+{1\over p_{34}^2}\Big)\Big({-\ell_R\cdot
p_{12}\over p_{12}^2(\ell_R\cdot p_1)}+{-\ell_R\cdot p_{12}\over
p_{12}^2(\ell_R\cdot p_2)}+{1\over p_{12}^2}\Big)\nonumber\\
&=&\int d^4\ell_R
~\delta^{+}(\ell_R^2)\delta^{+}((\ell_R+p_{12})^2){1\over2}\Big(-{1\over
2\ell_R\cdot p_3}-{1\over 2\ell_R\cdot p_4}+{1\over
p_{34}^2}\Big)\Big({1\over 2\ell_R\cdot p_1}+{1\over 2\ell_R\cdot
p_2}+{1\over p_{12}^2}\Big)~,~~~\eea
} \\ where in the second line we have used the on-shell conditions
$2\ell_L\cdot p_{12}=p_{12}^2$ and $2\ell_L\cdot p_{34}=p_{34}^2$
for $\Delta^{\mathcal{Q}}_{s_{12}}[\ell_L]$ and
$\Delta^{\mathcal{Q}}_{s_{12}}[\ell_R]$ respectively.

The contribution of $s_{12}$-cut channel by sewing two on-shell tree
amplitudes, after inserting the explicit expressions of
4-dimensional 4-point tree amplitudes, is given by
\bea \Delta^{\mathcal{F}}_{s_{12}}&=&{1\over 2}\int d^4\ell_L
d^4\ell_R
~\delta^4(\ell_R-\ell_L+p_{12})\delta^{+}(\ell_L^2)\delta^{+}(\ell_R^2)\nonumber\\
&&~~~~~~~~~~~~~~~~\times \Big({1\over (\ell_L-p_1)^2}+{1\over
(\ell_L-p_2)^2}+{1\over p_{12}^2}\Big)\Big({1\over
(\ell_R-p_3)^2}+{1\over (\ell_R-p_4)^2}+{1\over
p_{34}^2}\Big)~.~~~\eea
Note that the prefactor $1/2$ should be included for the {\sl double
unitarity cut} to avoid over-counting terms. Hence we
have
\bea \Delta^{\mathcal{F}}_{s_{12}}[\ell_L]&=&{1\over 2}\int
d^4\ell_L
~\delta^{+}(\ell_L^2)\delta^{+}((\ell_L-p_{12})^2)\nonumber\\
&&~~~~~~~~~~~~~~~\times\Big({1\over (\ell_L-p_1)^2}+{1\over
(\ell_L-p_2)^2}+{1\over p_{12}^2}\Big)\Big({1\over
(\ell_L+p_4)^2}+{1\over (\ell_L+p_3)^2}+{1\over
p_{34}^2}\Big)~,~~~\eea
as well as
\bea \Delta^{\mathcal{F}}_{s_{12}}[\ell_R]&=&{1\over 2}\int
d^4\ell_R
~\delta^{+}(\ell_R^2)\delta^{+}((\ell_R+p_{12})^2)\nonumber\\
&&~~~~~~~~~~~~~~~~\times \Big({1\over (\ell_R+p_2)^2}+{1\over
(\ell_R+p_1)^2}+{1\over p_{12}^2}\Big)\Big({1\over
(\ell_R-p_3)^2}+{1\over (\ell_R-p_4)^2}+{1\over
p_{34}^2}\Big)~.~~~\eea
Recall the on-shell cut conditions $\ell_L^2=0$, $\ell_R^2=0$ as
well as the massless conditions $p_i^2=0$, we immediately find
$\Delta^{\mathcal{Q}}_{s_{12}}[\ell_L]\simeq\Delta^{\mathcal{F}}_{s_{12}}[\ell_L]$
and
$\Delta^{\mathcal{Q}}_{s_{12}}[\ell_R]\simeq\Delta^{\mathcal{F}}_{s_{12}}[\ell_R]$.

Discussion of $s_{13}$- and $s_{14}$-cut channels is entirely
parallel to $s_{12}$-cut channel, where we only need to
replace $\{p_1,p_2,p_3,p_4\}\to \{p_1,p_3,p_2,p_4\}$ for
$s_{13}$ channel or $\{p_1,p_2,p_3,p_4\}\to \{p_4,p_1,p_2,p_3\}$ for
$s_{14}$ channel. This trivially verifies the equivalence
$\Delta^{\mathcal{Q}}_{s_{13}}[\ell_L]\simeq\Delta^{\mathcal{F}}_{s_{13}}[\ell_L]$,
$\Delta^{\mathcal{Q}}_{s_{13}}[\ell_R]\simeq\Delta^{\mathcal{F}}_{s_{13}}[\ell_R]$
and
$\Delta^{\mathcal{Q}}_{s_{14}}[\ell_L]\simeq\Delta^{\mathcal{F}}_{s_{14}}[\ell_L]$,
$\Delta^{\mathcal{Q}}_{s_{14}}[\ell_R]\simeq\Delta^{\mathcal{F}}_{s_{14}}[\ell_R]$.
Thus the equivalence $\cI^{\mathcal{Q}}\simeq \cI^{\mathcal{F}}$ is
confirmed by that for each cut channel.

%%%%%%%%%%%%%%%%%
\section{Applications in the Yang-Mills theory}
\label{secYangMills}
%%%%%%%%%%%%%%%%%%

In the previous section we illustrate the summation over all
$\mathcal{Q}$-cuts with amplitudes of scalar field theories. In
this section, we will clarify the summation over helicity states of
internal particles with color-ordered amplitudes of the Yang-Mills
theory. The convention adopted here is the t'Hooft-Veltman (HV)
regularization scheme \cite{tHooft:1972fi}, i.e., the loop momentum
will be $(4-2\epsilon)$-dimensional (as well as the polarization
vector) while the external momenta will be kept in the exact
4-dimension. More specifically, in the ${\cal Q}$-cut
construction (\ref{q-cut}), $\widetilde{\ell}_L$ and
$\widetilde{\ell}_R$ have non-vanishing components in the
$(-2\epsilon)$-dimension. As mentioned before, under dimensional
deformation, $\WH \ell_L$ and $\WH \ell_R$ will live in general
$D$-dimension, so the on-shell tree amplitudes used in
\eref{q-cut} should be the ones of $D$-dimension. These
amplitudes can be computed by Feynman rules, or expressions
generated by tree-level CHY
formula \cite{Cachazo:2013gna,Cachazo:2013hca, Cachazo:2013iea,
Cachazo:2014nsa,Cachazo:2014xea}, or $D$-dimensional BCFW recursion
relation \cite{Britto:2004ap,Britto:2005fq}. In this paper, we will
only focus on the 4-point one-loop amplitudes, and the necessary
4-point pure gluon tree amplitudes under various combinations of
helicity states\footnote{The helicity choices of $D$-dimensional
polarization vectors can be found in appendix \ref{convention}.}
$+,-,S_{A}$ are summarized in appendix \ref{4dimTree} for
reference. Furthermore, as emphasized in \S \ref{secReview},
while one can handle $\WH \ell_L$ and $\WH \ell_R$ in
$D$-dimension, when summing over physical helicity states in
\eref{q-cut}, we should restrict to $(4-2\epsilon)$-dimension in order to
produce the correct result.

%%%%%%%%%%%%%%%%%
\subsection{Color-ordered 4-point gluon amplitude
$\cA^{\oneloop}_4(1^+,2^+,3^+,4^+)$} \label{subsecYangMills1}
%%%%%%%%%%%%%%%%%%%

\begin{figure}
  \centering
  \includegraphics[width=6.5in]{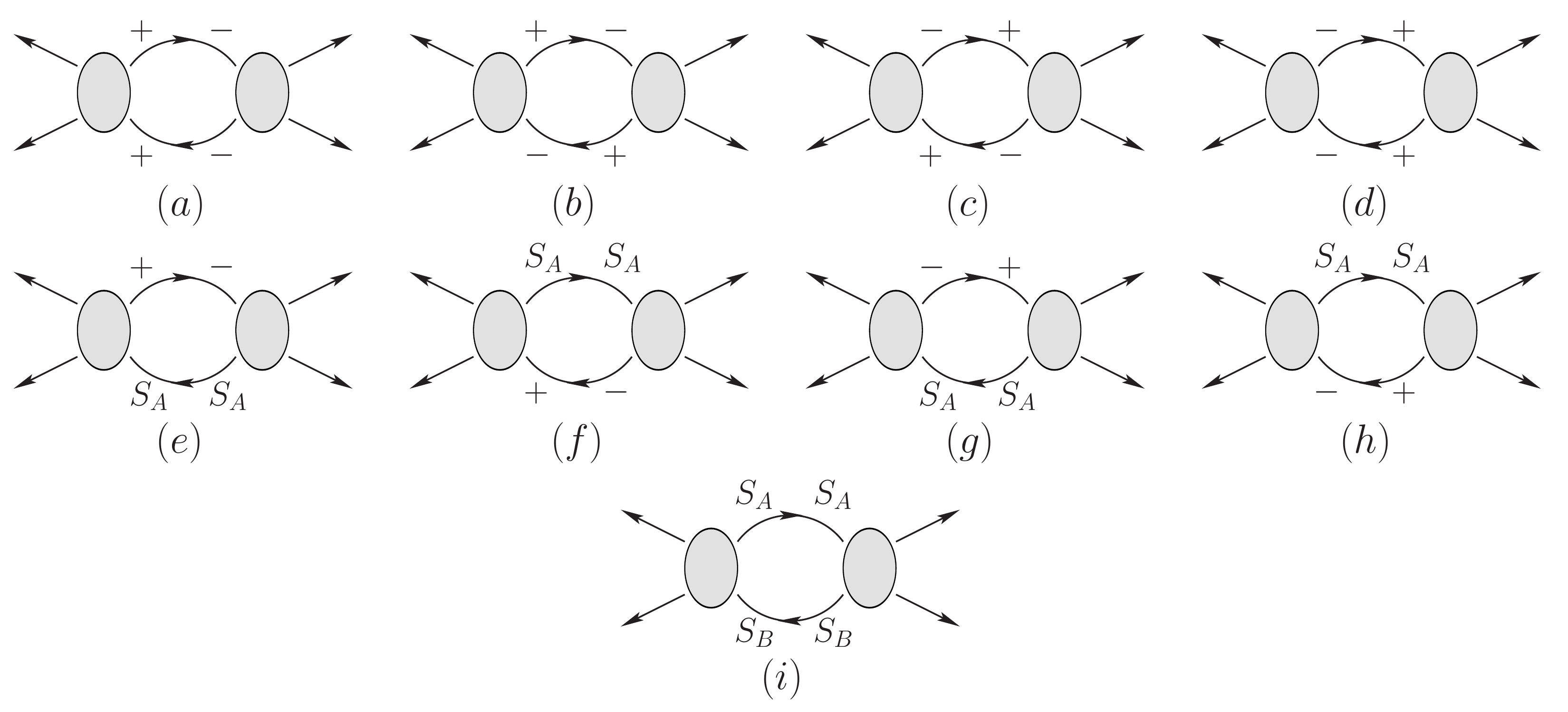}\\
  \caption{All possible internal states for the helicity sum of a
  given $\mathcal{Q}$-cut term, while the internal particles are $D$-dimensional gluons of $+,-$ and $S_A$ $(A=1,\ldots, \dim[\mu])$
  physical polarizations.
  }\label{Hsum}
\end{figure}

Let us start with a relatively simple example, namely the 4-point gluon
amplitude with all plus helicities
$A_4^{\oneloop}(1^+,2^+,3^+,4^+)$. For Yang-Mills theory, we will
not compute one-loop integrands by using Feynman rules, but directly
from ${\cal Q}$-cut construction and confirm its validity by the
cross-check of the unitarity cut method.

According to (\ref{q-cut}), this loop integrand is given by
\bea
\cI^{\mathcal{Q}}(\widetilde{\ell})=\sum_{h_1,h_2}\cA_{L}(1^+,2^+,
\widehat{\ell}_R^{~h_1},-\widehat{\ell}_L^{~h_2}){1\over
\widetilde{\ell}^2}{1\over (-2\widetilde{\ell}\cdot
p_{12}+p_{12}^2)}\cA_R(\widehat{\ell}_L^{~\bar{h}_2},-\widehat{\ell}_R^{~\bar{h}_1},3^+,4^+)+\cyclic{p_1,p_2,p_3,p_4}~,~~~\eea
which contains four $\mathcal{Q}$-cut terms, and the helicities
$h_1,h_2$ of internal states should be summed in line with the nine
cases shown in Figure (\ref{Hsum}). If the internal gluons are
4-dimensional, only the first four diagrams in Figure
(\ref{Hsum}) may exist, which all vanish due to
$A^{\tree}_4(+,+,+,+)=0$, $A^{\tree}_4(+,+,+,-)=0$ in
4-dimension. When the loop momentum is taken to be
$D$-dimensional and all external momenta have positive helicities, the
cases of Figure (\ref{Hsum}.a), (\ref{Hsum}.d), (\ref{Hsum}.e),
(\ref{Hsum}.f), (\ref{Hsum}.g), (\ref{Hsum}.h) have no
contribution due to the vanishing tree amplitudes. So we only
need to sum over Figure (\ref{Hsum}.b), (\ref{Hsum}.c) and
(\ref{Hsum}.i) with $S_A=S_B$, given by
\bea \cI^{\mathcal{Q}}(\widetilde{\ell})&=&\cA_{L}(1^+,2^+,
\widehat{\ell}_R^{~+},-\widehat{\ell}_L^{~-}){1\over
\widetilde{\ell}^2}{1\over (-2\widetilde{\ell}\cdot
p_{12}+p_{12}^2)}\cA_R(\widehat{\ell}_L^{~+},-\widehat{\ell}_R^{~-},3^+,4^+)\nonumber\\
&&~~~~+\cA_{L}(1^+,2^+,
\widehat{\ell}_R^{~-},-\widehat{\ell}_L^{~+}){1\over
\widetilde{\ell}^2}{1\over (-2\widetilde{\ell}\cdot
p_{12}+p_{12}^2)}\cA_R(\widehat{\ell}_L^{~-},-\widehat{\ell}_R^{~+},3^+,4^+)\nonumber\\
&&~~~~~~~~+\sum_{A}^{\dim[\mu]}\cA_{L}(1^+,2^+,
\widehat{\ell}_R^{~S_A},-\widehat{\ell}_L^{~S_A}){1\over
\widetilde{\ell}^2}{1\over (-2\widetilde{\ell}\cdot
p_{12}+p_{12}^2)}\cA_R(\widehat{\ell}_L^{~S_A},-\widehat{\ell}_R^{~S_A},3^+,4^+)\nonumber\\
&&~~~~~~~~~~~~~~~~+\cyclic{p_1,p_2,p_3,p_4}~.~~~\eea
where again $\widehat{\ell}_L=\alpha_L(\widetilde{\ell}+\eta)$ and
$\widetilde{\ell}=\ell+\mu$.

The $D$-dimensional tree amplitudes, as given in appendix
\ref{4dimTree}, depend on the reference momentum $q$. However for
this example, the product $\cA_L\cA_R$ in each term is independent
of $q$, thus the loop integrand is also independent of $q$, which
serves as a consistency check. More explicitly, let us first write
the general $D$-dimensional vector as $\widehat{\ell}_L\equiv
(\widetilde{\ell},\eta)=(\ell,\mu,\eta)$, we have
\bea \cA_{L}(1^+,2^+,
\widehat{\ell}_R^{~+},-\widehat{\ell}_L^{~-})\cA_R(\widehat{\ell}_L^{~+},-\widehat{\ell}_R^{~-},3^+,4^+)={\spbb{1~2}\spbb{3~4}\over
\spaa{1~2}\spaa{3~4}}{(\mu^2+\eta^2)^2\over (2\widetilde{\ell}\cdot
p_1)(2\widetilde{\ell}\cdot p_4)}~,~~~\eea
Under the massless conditions of $\WH \ell_L, \WH\ell_R$, we should
make the following replacement $\eta^2\to \widetilde{\ell}^2$ as
well as $\widetilde{\ell}\to \alpha_L\widetilde{\ell}$ and
$\mu\to\alpha_L\mu$ with $\alpha_L=p_{12}^2/(2\widetilde{\ell}\cdot
p_{12})$ in succession, and find
\bea \cA_{L}(1^+,2^+,
\widehat{\ell}_R^{~+},-\widehat{\ell}_L^{~-})\cA_R(\widehat{\ell}_L^{~+},-\widehat{\ell}_R^{~-},3^+,4^+)={\spbb{1~2}\spbb{3~4}\over
\spaa{1~2}\spaa{3~4}}{(\mu^2+\widetilde{\ell}^2)^2(p_{12}^2/(2\widetilde{\ell}\cdot
p_{12}))^2\over (2\widetilde{\ell}\cdot p_1)(2\widetilde{\ell}\cdot
p_4)}~.~~~\eea
The rest two diagrams $$\cA_{L}(1^+,2^+,
\widehat{\ell}_R^{~-},-\widehat{\ell}_L^{~+})\cA_R(\widehat{\ell}_L^{~-},-\widehat{\ell}_R^{~+},3^+,4^+)~~\mbox{and}~~\cA_{L}(1^+,2^+,
\widehat{\ell}_R^{~S_A},-\widehat{\ell}_L^{~S_A})\cA_R(\widehat{\ell}_L^{~S_A},-\widehat{\ell}_R^{~S_A},3^+,4^+)$$
lead to exactly the same results. Hence we get  the  integrand in
$\mathcal{Q}$-cut representation as
\bea
\cI^{\mathcal{Q}}(\widetilde{\ell})&=&(2-2\epsilon){\spbb{1~2}\spbb{3~4}\over
\spaa{1~2}\spaa{3~4}}{(\mu^2+\widetilde{\ell}^2)^2(p_{12}^2/(2\widetilde{\ell}\cdot
p_{12}))^2\over \widetilde{\ell}^2(-2\widetilde{\ell}\cdot
p_{12}+p_{12}^2)(2\widetilde{\ell}\cdot p_1)(2\widetilde{\ell}\cdot
p_4)}+\cyclic{p_1,p_2,p_3,p_4}~,~~~\label{YM-4posi-Q} \eea
where we have summed over helicity states in
$(4-2\eps)$-dimension (especially including the $S_A$ components in
$\dim[\mu]=(-2\epsilon)$-dimension).

With this result, we can use the unitarity cut method to do the cross-check.
Consider first the $s_{12}$-cut channel, we
need to evaluate
\bea \Delta^{\mathcal{Q}}_{s_{12}}=\int d^{4-2\epsilon}\wt\ell_L
d^{4-2\epsilon}\wt\ell_R
~\delta^{4-2\epsilon}(\wt\ell_R-\wt\ell_L+p_{12})\delta^{+}(\wt\ell_L^{~2})\delta^{+}(\wt\ell_R^{~2})\times
\cI^{\mathcal{Q}}(\wt\ell)~\wt\ell_L^{~2}\wt\ell_R^{~2}~.~~~\eea
Again we have two different choices of identifications. If we identify
$\W\ell=\W\ell_L$ and integrate over $\wt\ell_R$ against momentum
conservation, the factor $\wt\ell_L^{2}\wt\ell_R^{2}$ becomes
$\wt\ell_L^{2}(\wt\ell_L-p_{12})^{2}\to
\wt\ell_L^{2}(-2\wt\ell_L\cdot p_{12}+p_{12}^2)$. Upon the
cut conditions $\wt\ell_L^{2}=0$, $\wt\ell_R^{2}=0$, it trivially
vanishes. Thus this picks out the terms with denominator
$\wt\ell^{2}(-2\wt\ell\cdot p_{12}+p_{12}^2)$ in \eref{YM-4posi-Q}
while all other terms vanish. Hence
\bea \Delta^{\mathcal{Q}}_{s_{12}}[\wt\ell_L]=\int
d^{4-2\epsilon}\wt\ell_L
\delta^{+}(\wt\ell_L^{~2})\delta^{+}((\wt\ell_L-p_{12})^{2}){\spbb{1~2}\spbb{3~4}\over
\spaa{1~2}\spaa{3~4}}{(2-2\epsilon)\mu^4\over
(2\widetilde{\ell}_L\cdot p_1)(2\widetilde{\ell}_L\cdot
p_4)}~.~~~\eea
If we identify $\W\ell=\W\ell_R$ and integrate over $\wt\ell_L$ against momentum
conservation, only the terms with denominator
$\wt\ell^{2}(-2\wt\ell\cdot p_{34}+p_{34}^2)$ can survive.
Thus we have
\bea \Delta^{\mathcal{Q}}_{s_{12}}[\wt\ell_R]=\int
d^{4-2\epsilon}\wt\ell_R
~\delta^{+}(\wt\ell_R^{~2})\delta^{+}((\wt\ell_R+p_{12})^{2}){\spbb{3~4}\spbb{1~2}\over
\spaa{3~4}\spaa{1~2}}{(2-2\epsilon)\mu^4\over
(2\widetilde{\ell}_R\cdot p_3)(2\widetilde{\ell}_R\cdot
p_2)}~.~~~\eea
For the $s_{12}$-channel, the standard unitarity cut
method yields
\bea \Delta^{\mathcal{F}}_{s_{12}}&=&\sum_{h_1,h_2}\int
d^{4-2\epsilon}\wt\ell_L d^{4-2\epsilon}\wt\ell_R
~\delta^{4-2\epsilon}(\wt\ell_R-\wt\ell_L+p_{12})\delta^{+}(\wt\ell_L^{~2})\delta^{+}(\wt\ell_R^{~2})\nonumber\\
&&~~~~~~~~~~~~~~~~~~~~~~~~~~~~~~~~~~~~~~~~\times
\cA_L(1,2,\wt\ell_R^{~h_1},-\wt\ell_L^{~h_2})\cA_R(\wt\ell_L^{~\bar{h}_2},-\wt\ell_R^{~\bar{h}_1},3,4)~,~~~\eea
where again the tree amplitudes $\cA_{L}, \cA_R$ are
$(4-2\epsilon)$-dimensional and the helicity states should be
summed over the nine diagrams in Figure (\ref{Hsum}). Inserting the results
in appendix \ref{4dimTree}, we get
\bea \Delta^{\mathcal{F}}_{s_{12}}&=&\int d^{4-2\epsilon}\wt\ell_L
d^{4-2\epsilon}\wt\ell_R
~\delta^{4-2\epsilon}(\wt\ell_R-\wt\ell_L+p_{12})\delta^{+}(\wt\ell_L^{~2})\delta^{+}(\wt\ell_R^{~2}){\spbb{1~2}\spbb{3~4}\over\spaa{1~2}\spaa{3~4}}{-(2-2\epsilon)\mu^4\over(\wt\ell_L-p_1)^2(\wt\ell_R-p_3)^2}~.~~~\eea
Depending on the integration order of $\W\ell_R$ and $\W\ell_L$,
it gives
\bea \Delta^{\mathcal{F}}_{s_{12}}[\wt\ell_L]&=&\int
d^{4-2\epsilon}\wt\ell_L
~\delta^{+}(\wt\ell_L^{~2})\delta^{+}((\wt\ell_L-p_{12})^2){\spbb{1~2}\spbb{3~4}\over\spaa{1~2}\spaa{3~4}}{-(2-2\epsilon)\mu^4\over(\wt\ell_L-p_1)^2(\wt\ell_L+p_4)^2}~,~~~\eea
or
\bea \Delta^{\mathcal{F}}_{s_{12}}[\wt\ell_R]&=&\int
d^{4-2\epsilon}\wt\ell_R
~\delta^{+}(\wt\ell_R^{~2})\delta^{+}((\wt\ell_R+p_{12})^2){\spbb{1~2}\spbb{3~4}\over\spaa{1~2}\spaa{3~4}}{-(2-2\epsilon)\mu^4\over(\wt\ell_R+p_2)^2(\wt\ell_R-p_3)^2}~.~~~\eea
Upon the on-shell conditions $\wt\ell_L^2=0$, $\wt\ell_R^2=0$ and
$p_i^2=0$, we have confirmed the equivalence
$\Delta^{\mathcal{Q}}_{s_{12}}\simeq \Delta^{\mathcal{F}}_{s_{12}}$.

Since all external particles have positive helicities, the cross-check for
$s_{14}$-channel can be easily done by shifting $p_i\to
p_{i+1}$ in the previous computation.

%%%%%%%%%%%%%%%%%
\subsection{Color-ordered 4-point gluon amplitude
$\cA^{\oneloop}_4(1^+,2^+,3^+,4^-)$} \label{subsecYangMills2}
%%%%%%%%%%%%%%%%%%%

Now let us consider a more complicated example
$\cA^{\oneloop}_4(1^+,2^+,3^+,4^-)$. Its one-loop integrand produced
by $\mathcal{Q}$-cut construction is
\bea
\cI^{\mathcal{Q}}(\widetilde{\ell})=\sum_{h_1,h_2}\cA_{L}(1^+,2^+,
\widehat{\ell}_R^{~h_1},-\widehat{\ell}_L^{~h_2}){1\over
\widetilde{\ell}^2}{1\over (-2\widetilde{\ell}\cdot
p_{12}+p_{12}^2)}\cA_R(\widehat{\ell}_L^{~\bar{h}_2},-\widehat{\ell}_R^{~\bar{h}_1},3^+,4^-)+\cyclic{p_1,p_2,p_3,p_4}~.~~~\eea
The $D$-dimensional tree amplitudes can be referred in appendix
\ref{4dimTree}. We still need to sum over the nine diagrams in
Figure (\ref{Hsum}). But those of Figure (\ref{Hsum}.a),
(\ref{Hsum}.e), (\ref{Hsum}.f) have no contribution due to the
vanishing tree amplitudes, while that of Figure (\ref{Hsum}.i)
only contributes when $S_A=S_B$. Hence we have, for example,{\small
\bea
&&\sum_{h_1,h_2}\cA_L(1^+,2^+,\widehat{\ell}_R^{~h_1},-\WH\ell_L^{~h_2})\cA_R(\widehat{\ell}_L^{~\bar{h}_2},-\widehat{\ell}_R^{~\bar{h}_1},3^+,4^-)\\
&=&\sum_{A=1}^{\dim[\mu]}\Big(\cA_L(1^+,2^+,\widehat{\ell}_R^{~S_A},-\widehat{\ell}_L^{~S_A})\cA_R(\widehat{\ell}_L^{~S_A},-\widehat{\ell}_R^{~S_A},3^+,4^-)\Big)+\sum_{A=1}^{\dim[\mu]} \Big( \cA_L(1^+,2^+,\widehat{\ell}_R^{~-},-\widehat{\ell}_L^{~S_A})\cA_R(\widehat{\ell}_L^{~S_A},-\widehat{\ell}_R^{~+},3^+,4^-)\Big)\nonumber\\
&&+\sum_{A=1}^{\dim[\mu]}\Big(\cA_L(1^+,2^+,\widehat{\ell}_R^{~S_A},-\widehat{\ell}_L^{~-})\cA_R(\widehat{\ell}_L^{~+},-\widehat{\ell}_R^{~S_A},3^+,4^-)\Big)+\cA_L(1^+,2^+,\widehat{\ell}_R^{~-},-\widehat{\ell}_L^{~-})\cA_R(\widehat{\ell}_L^{~+},-\widehat{\ell}_R^{~+},3^+,4^-)\nonumber\\
&&+\cA_L(1^+,2^+,\widehat{\ell}_R^{~-},-\widehat{\ell}_L^{~+})\cA_R(\widehat{\ell}_L^{~-},-\widehat{\ell}_R^{~+},3^+,4^-)+\cA_L(1^+,2^+,\widehat{\ell}_R^{~+},-\widehat{\ell}_L^{~-})\cA_R(\widehat{\ell}_L^{~+},-\widehat{\ell}_R^{~-},3^+,4^-)~.~~~\nonumber\eea
} \\ Different from the previous example, here each term
$\cA_L\cA_R$ depends on the reference momentum $q$, which is the
gauge choice for the polarization vector.
%means that for the
%helicity choice of polarization vector defined in Appendix
%(\ref{convention}), the result of each $\cA_L\cA_R$ depends on the gauge choice.
However, after summing all terms we do get a
$q$-independent result, namely
\bea
\sum_{h_1,h_2}\cA_L(1^+,2^+,\widehat{\ell}_R^{~h_1},-\WH\ell_L^{~h_2})\cA_R(\widehat{\ell}_L^{~\bar{h}_2},-\widehat{\ell}_R^{~\bar{h}_1},3^+,4^-)=(2-2\epsilon)\frac{(\mu^2+\eta^2)
\spbb{2~1} \spab{4|{\ell}_L|3}^2}{\spaa{1~2} \spaa{3~4}
\spbb{4~3}(2\WH\ell_L\cdot p_1)(2\WH\ell_L\cdot
p_4)}~.~~~\label{HsumPPPM1}\eea
The detailed computation of helicity sum can be referred in appendix
\ref{HsumDetail}. This result is rather compact, yet it enjoys the
advantage of the spinor-helicity formalism. Another expression in terms of the
standard Mandelstam variables can be obtained after tensor reduction,
given by
\bea &&(2-2\epsilon)(\mu^2+\eta^2)
{\spbb{1~3}\spab{4|3|2}\over\spaa{1~3}\spab{2|3|4}}\Big({s_{14}-s_{13}\over
4 s_{14} s_{13}}+{s_{12}s_{14}+s_{12}( \WH\ell_L\cdot
p_1)+2i\epsilon(\WH\ell_L,1,2,3)\over 8s_{14}s_{13}(
\WH\ell_L\cdot p_4)}~~~~~\label{HsumPPPM2}\\
&&+{s_{12}s_{14}+2s_{13}(\mu^2+\eta^2)+4i\epsilon(\WH\ell_L,1,2,3)\over
32 s_{13}(\WH\ell_L\cdot p_1)( \WH\ell_L\cdot
p_4)}+{s_{12}s_{14}^2+(s_{14}^2+s_{13}^2)( \WH\ell_L\cdot
p_4)-2(s_{13}-s_{14})i\epsilon(\WH\ell_L,1,2,3)\over
8s_{12}s_{14}s_{13}(\WH\ell_L\cdot p_1)}\Big)~,~~~\nonumber\eea
where $s_{ij}\equiv p_{ij}^2$. To finally reach the ${\cal Q}$-cut
representation, we can impose $\eta^2\to\wt\ell^2$, $\wt\ell\to
\alpha_L\wt\ell$, $\mu\to \alpha_L\mu$ and
$\alpha_L=p_{12}^2/(2\ell_L\cdot p_{12})$ for either
(\ref{HsumPPPM1}) or (\ref{HsumPPPM2}). For simplicity, we choose the
expression (\ref{HsumPPPM1}) and get
\bea
\cI^{\mathcal{Q}}(\widetilde{\ell})&=&{(2-2\epsilon)\spab{4|\ell|3}^2(\mu^2+\wt\ell^2)(p_{12}^2/(2\wt\ell\cdot
p_{12})^2)\over
\spaa{1~2}^2\widetilde{\ell}^2(-2\widetilde{\ell}\cdot
p_{12}+p_{12}^2)(2\wt\ell\cdot p_1)(2\wt\ell\cdot
p_4)}+{(2-2\epsilon)\spab{4|\ell|1}^2(\mu^2+\wt\ell^2)(p_{23}^2/(2\wt\ell\cdot
p_{23})^2)\over
\spaa{2~3}^2\widetilde{\ell}^2(-2\widetilde{\ell}\cdot
p_{23}+p_{23}^2)(2\wt\ell\cdot p_2)(2\wt\ell\cdot
p_1)}\nonumber\\
&&+{(2-2\epsilon)\spab{4|\ell|3}^2(\mu^2+\wt\ell^2)(p_{34}^2/(2\wt\ell\cdot
p_{34})^2)\over
\spaa{1~2}^2\widetilde{\ell}^2(-2\widetilde{\ell}\cdot
p_{34}+p_{34}^2)(2\wt\ell\cdot p_3)(2\wt\ell\cdot
p_2)}+{(2-2\epsilon)\spab{4|\ell|1}^2(\mu^2+\wt\ell^2)(p_{41}^2/(2\wt\ell\cdot
p_{41})^2)\over
\spaa{2~3}^2\widetilde{\ell}^2(-2\widetilde{\ell}\cdot
p_{41}+p_{41}^2)(2\wt\ell\cdot p_4)(2\wt\ell\cdot
p_3)}~.~~~\label{YMPPPMQcut}\eea
In fact, by using identities
\bea \Big({p_{ij}^2\over 2\widetilde{\ell}\cdot
p_{ij}}\Big)^2={(p_{ij}^2-2\widetilde{\ell}\cdot
p_{ij}+2\widetilde{\ell}\cdot p_{ij})^2\over (2\widetilde{\ell}\cdot
p_{ij})^2}={(p_{ij}^2-2\widetilde{\ell}\cdot p_{ij})^2\over
(2\widetilde{\ell}\cdot
p_{ij})^2}+{2(p_{ij}^2-2\widetilde{\ell}\cdot p_{ij})\over
(2\widetilde{\ell}\cdot p_{ij})}+1~,~~~\label{scalefreeIden}\eea
the factors $(p_{ij}^2/(2\wt\ell\cdot p_{ij}))^2$ can be effectively
replaced by 1 after dropping scale-free terms, which simplifies the
result into
\bea
\cI^{\mathcal{Q}}(\widetilde{\ell})&=&{(2-2\epsilon)\spab{4|\ell|3}^2(\mu^2+\wt\ell^2)\over
\spaa{1~2}^2\widetilde{\ell}^2(-2\widetilde{\ell}\cdot
p_{12}+p_{12}^2)(2\wt\ell\cdot p_1)(2\wt\ell\cdot
p_4)}+{(2-2\epsilon)\spab{4|\ell|1}^2(\mu^2+\wt\ell^2)\over
\spaa{2~3}^2\widetilde{\ell}^2(-2\widetilde{\ell}\cdot
p_{23}+p_{23}^2)(2\wt\ell\cdot p_2)(2\wt\ell\cdot
p_1)}\nonumber\\
&&+{(2-2\epsilon)\spab{4|\ell|3}^2(\mu^2+\wt\ell^2)\over
\spaa{1~2}^2\widetilde{\ell}^2(-2\widetilde{\ell}\cdot
p_{34}+p_{34}^2)(2\wt\ell\cdot p_3)(2\wt\ell\cdot
p_2)}+{(2-2\epsilon)\spab{4|\ell|1}^2(\mu^2+\wt\ell^2)\over
\spaa{2~3}^2\widetilde{\ell}^2(-2\widetilde{\ell}\cdot
p_{41}+p_{41}^2)(2\wt\ell\cdot p_4)(2\wt\ell\cdot
p_3)}~.~~~\label{YMPPPMQcut-1}\eea
Now we use the unitarity cut method to do the cross-check for this result,
followed by the same strategy in the previous
examples. For the $s_{12}$ channel, the computation of
\bea \Delta^{\mathcal{Q}}_{s_{12}}=\int d^{4-2\epsilon}\wt\ell_L
d^{4-2\epsilon}\wt\ell_R
~\delta^{4-2\epsilon}(\wt\ell_R-\wt\ell_L+p_{12})\delta^{+}(\wt\ell_L^{~2})\delta^{+}(\wt\ell_R^{~2})\times
\cI^{\mathcal{Q}}(\wt\ell)~\wt\ell_L^{~2}\wt\ell_R^{~2}~~~~\eea
can be done by either identifying $\W\ell=\W\ell_L$ and integrating
over $\wt\ell_R$, or identifying $\W\ell=\W\ell_R$ and integrating over
$\wt\ell_L$. For the former choice, the first term in
(\ref{YMPPPMQcut-1}) is picked out, and the on-shell conditions
$\wt\ell_L^2=0$, $\wt\ell_R^2=0$ imply $p_{12}^2/(2\wt\ell\cdot
p_{12})=1$,  hence
\bea \Delta^{\mathcal{Q}}_{s_{12}}[\wt\ell_L]=\int
d^{4-2\epsilon}\wt\ell_L
\delta^{+}(\wt\ell_L^{~2})\delta^{+}((\wt\ell_L-p_{12})^{2}){\spab{4|\ell_L|3}^2\over\spaa{1~2}^2}{(2-2\epsilon)\mu^2\over(2\wt\ell_L\cdot
p_1)(2\wt\ell_L\cdot p_4)}~.~~~\eea
For the latter one, we have
\bea \Delta^{\mathcal{Q}}_{s_{12}}[\wt\ell_R]=\int
d^{4-2\epsilon}\wt\ell_R
~\delta^{+}(\wt\ell_R^{~2})\delta^{+}((\wt\ell_R+p_{12})^{2}){\spab{4|\ell_L|3}^2\over\spaa{1~2}^2}{(2-2\epsilon)\mu^2\over(2\wt\ell_R\cdot
p_3)(2\wt\ell_R\cdot p_2)}~.~~~\eea
The traditional unitarity cut method gives the following result by
sewing two $(4-2\epsilon)$-dimensional tree amplitudes, which is
\bea \Delta^{\mathcal{F}}_{s_{12}}&=&\int d^{4-2\epsilon}\wt\ell_L
d^{4-2\epsilon}\wt\ell_R
~\delta^{4-2\epsilon}(\wt\ell_R-\wt\ell_L+p_{12})\delta^{+}(\wt\ell_L^{~2})\delta^{+}
(\wt\ell_R^{~2}){\spab{4|\ell_L|3}^2\over\spaa{1~2}^2}{-(2-2\epsilon)\mu^4\over(\wt\ell_L-p_1)^2
(\wt\ell_R-p_3)^2}~.~~~\eea
Just like the computation in the previous subsection, depending on the
integration order of $\wt\ell_L$ and $\wt\ell_R$, it can be
immediately shown that $\Delta_{s_{12}}^{\mathcal{Q}}\simeq
\Delta_{s_{12}}^{\mathcal{F}}$. Similar computation can be done for
$s_{14}$-channel. Since for each unitarity cut the
equivalence is valid, the equivalence between the loop integrands of ${\cal
Q}$-cut representation and traditional Feynman diagrams has
been confirmed.

%%%%%%%%%%%%%%%%%
\section{Conclusion}
\label{secConclusion}
%%%%%%%%%%%%%%%%%%

In this paper, we take the integrands of one-loop amplitudes in
scalar field and Yang-Mills theories as examples to elaborate
various aspects of the newly proposed ${\cal Q}$-cut construction,
specifically the summation over distinct $\mathcal{Q}$-cuts, as well as
the internal helicity states. Furthermore, a cross-check
using traditional unitarity cut has been provided, establishing
the connection between this new algorithm and
traditional computational techniques for loop amplitudes.

The construction of one-loop integrands has become a popular topic
recently, by generalizing the tree-level massless on-shell CHY
formulation \cite{Cachazo:2013hca,Cachazo:2013iea} to loop-level
\cite{Geyer:2015bja}, using the ambitwistor string theory (see also
\cite{Baadsgaard:2015hia, He:2015yua}).
%In \cite{He:2015yua}, the author construct the one-loop integrand in bi-adjoint scalar theory. Big advances have been taken by considering the scattering equations in ambitwistor string theories\cite{Geyer:2015bja,Geyer:2015jch}.
Furthermore, the authors of ref. \cite{Geyer:2015jch} showed that their result was in
fact equivalent to the $\mathcal{Q}$-cut representation, which is an
affirmative support of this approach.

%In this paper, we showed in examples that the loop integrand in $\mathcal{Q}$-cut representation has one-to-one correspondence to the one produced by Feynman diagrams, it would be great if some generic proof(or mapping rules) can be found for the general case, where one can maps the terms of $\mathcal{Q}$-cut representation to those of Feynman diagrams with standard propagators.

As also pointed out in \cite{Baadsgaard:2015twa}, there are many
questions to be investigated. First of all, it will be extremely
significant to compute a non-trivial two-loop amplitude via the
$\mathcal{Q}$-cut construction, to test its ability and
potential advantages over other methods. Although the general
framework to handle two or more loops has been prescribed,
carrying out the particulars is unquestionably favorable. Secondly, the ideas encoded in
the ${\cal Q}$-cut construction, especially the use of linear
propagators, possibly opens a new window for the current researches
on the integrand-level reduction by computational algebraic geometry
method, or the integral-level reduction by improved IBP relations. It
is worth exploring whether these two multi-loop integrand reduction
techniques can be applied directly to the $\mathcal{Q}$-cut
representation. Finally, at this moment, the ${\cal Q}$-cut
construction is still restricted to massless theories, so it will be naturally
important to generalize it to massive theories.

%%%%%%%%%%%%%%%%%%%%%%%%%%%%%%%%%%%%%%%%%%%%%%%%%%%%%%%

\section*{Acknowledgments}

We would like to thank Emil Bjerrum-Bohr, Poul Henrik Damgaard and
Song He for reading the draft. BF would like to thank the
participants of workshop {\sl Dark Energy and Fundamental Theory},
supported by the Special Fund for Theoretical Physics from the
National Natural Science Foundation of China with Grant No.11447613,
for discussions. This work is supported by the National Natural
Science Foundation of China with Grant No.11135006, No.11125523 and
No.11575156. RH also would like to acknowledge the support by
Chinese Postdoctoral Administrative Committee.

%BF is grateful to the Qiu-Shi Fund and the Chinese NSF under
%contracts No.11031005, No.11135006 and No.11125523. RH would also
%like to thank the supporting from Chinese Postdoctoral
%Administrative Committee.

\appendix

%%%%%%%%%%%%%%%%%
\section{Helicity choices of polarization vectors}
\label{convention}
%%%%%%%%%%%%%%%%%%%

Throughout this paper, we use the $D$-dimensional metric
$\eta_{\mu\nu}=\mbox{diag}(+,-,-,\cdots,-)$. After cutting two
internal lines of a one-loop diagram, we shall get two on-shell tree
amplitudes, and each has two legs corresponding to the cut
propagators in $D$-dimension. To describe such amplitudes, we define
\bea \WH \ell=
(\ell,\vec{\mu})~~~,~~~\vec{\mu}=(\mu_1,\ldots,\mu_d)~,~~~\label{D-ell}\eea
where $\ell$ is the 4-dimensional component and $\vec{\mu}$ is a
vector in the extra $d$-dimension ($d=D-4$). For the Euclidean
$d$-dimensional space, we adopt the basis $e_A$ with $A=1,...,d$ such that its
components $\mu_{i}=\delta_{iA}$ for $i=1,...,d$. The massless
condition of $\WH \ell$ is then $\ell^2-\vec{\mu}^2=0$.

Now we define the polarization vectors along with the massless
momentum $\WH \ell$. The transverse condition $\eps\cdot \WH \ell=0$
tells that the transverse space is $(D-2)$-dimensional. However, due to the
gauge degree of freedom $\eps_i\sim \eps_i+\a \ell$, to completely fix
$\eps_i$ we need to introduce an auxiliary momentum $q$, which
corresponds to the gauge choice. For convenience, we will choose $q$
to be null in 4-dimension and $q\cdot \ell\neq 0$. With
this choice, we can fix transverse polarization vectors by imposing
the extra condition $\eps_i\cdot q=0$. According to the
principles above, firstly, we define the null 4-dimensional momentum as
\bea \ell^{\bot}=\ell-{\vec{\mu}^2\over
\Spab{q|\ell|q}}q~~~,~~~~({\ell^{\bot}})^2=0~,~~~\label{ell-bot}\eea
where $\ell$ is the 4-dimensional component of $\WH \ell$ (see
\eref{D-ell}). With the null momentum $\ell^{\bot}$, the transverse
polarization vectors can be defined as
\bea &&\epsilon^+(\widehat{\ell},q)=
\Big({\spab{q|\gamma_\nu|\ell^{\bot}}\over
\sqrt{2}\spaa{q~\ell^{\bot}}},\vec{0}_d\Big)~~~,~~~\epsilon^-(\widehat{\ell},q)=
\Big({\spab{\ell^{\bot}|\gamma_\nu|q}\over
\sqrt{2}\spbb{\ell^{\bot}~q}},\vec{0}_d\Big)~,~~~\nonumber\\
&&\epsilon^{S_A}(\widehat{\ell},q)=\Big({2 \mu_A\over
\spab{q|\ell|q}}q,e_A\Big)~~~,~~~A=1,...,d~,~~~\label{Pol-D-1}\eea
where $\vec{0}$ denotes the $d$-dimensional vanishing vector, $e_A$ is
the basis vector of the extra $d$-dimension and $\mu_A$ is the $A$-th
component of $\vec{\mu}$ in \eref{D-ell}. The reason of regarding
$S_A$ as a superscript is that these polarization states behave like
scalars with respect to the 4-dimension. The rest two
polarization vectors are longitudinal and time-like, defined as
\bea \epsilon^{L}(\widehat{\ell},q)=\WH
\ell~~~,~~~\epsilon^{T}(\widehat{\ell},q)=(q,\vec{0}_d)~.~~~\label{Pol-D-2}\eea
These $D$ polarization vectors possess the following relations
\bea &&\epsilon^{\pm}\cdot q=\epsilon^{\pm}\cdot
\ell=\epsilon^{\pm}\cdot
\epsilon^{\pm}=0~~,~~\epsilon^+\cdot\epsilon^-=-1~,~~~\nonumber\\
&&\epsilon^{L}\cdot
\epsilon^{T}=\spab{q|{\ell}|q}~~,~~\epsilon^{S_A}\cdot
\epsilon^{L,T}=0~~,~~\epsilon^{S_A}\epsilon^{S_B}=-\delta^{AB}~.~~~\label{inner-product}\eea
With this setup, we have the metric decomposition
\bea
\eta_{\mu\nu}={\epsilon^{L}\epsilon^{T}+\epsilon^{T}\epsilon^{L}\over\spab{q|{\ell}|q}}
-\epsilon^{+}\epsilon^{-}-\epsilon^{-}\epsilon^{+}-\sum_{A=1}^d\epsilon^{S_A}\epsilon^{S_A}~.~~~~~\label{metric-decom}\eea
The convention above is for general integer dimension $D$, but of
course it can be trivially generalized to $(4-2\epsilon)$-dimension,
for which we should replace $d\to -2\epsilon$.

%%%%%%%%%%%%%%%%%
\section{$D$-dimensional 4-point tree amplitudes of the Yang-Mills theory}
\label{4dimTree}
%%%%%%%%%%%%%%%%%%%

The computation of loop integrands by $\mathcal{Q}$-cut
representation requires on-shell tree amplitudes in general
$D$-dimension. There are several ways to get these
amplitudes. The first one is to use Feynman diagrams. The second
one is to use the CHY formula, which holds in arbitrary
dimensions \cite{Cachazo:2013hca,Cachazo:2013iea}. The third one is to
use the on-shell recursion
relation \cite{Britto:2004ap,Britto:2005fq}, starting from the
3-point seed amplitudes. Note that, for our formalism which
involves extra dimensions and parameters, it is no longer obvious to
get the 3-point amplitudes solely by little group scaling.
Therefore conservatively, we still start from listing the 3-point
amplitudes given by simple Feynman rules,
\bea
\cA(1,2,3)&=-i\sqrt{2}\,\big[(p_2\cdot\epsilon_3)(\epsilon_1\cdot\epsilon_2)
+(p_3\cdot\epsilon_1)(\epsilon_2\cdot\epsilon_3)+(p_1\cdot\epsilon_2)(\epsilon_3\cdot\epsilon_1)\big]~.~~~\eea
For physical polarizations $+,-,S_A$ $(A=1,2,\ldots, d)$, we need the
following 3-point tree amplitudes,
\bea
&&\cA_3(1^{+},2^{+},3^{+})=A(1^{-},2^{-},3^{-})=0~~,~~\cA_3(1^{+},2^{+},3^{-})=-{\spbb{1~2}^3\over
\spbb{2~3}\spbb{3~1}}~~,~~\cA_3(1^{-},2^{-},3^{+})={\spaa{1~2}^3\over\spaa{2~3}\spaa{3~1}}~,~~~\nonumber\\
&&\cA_3(1^{+},2^{+},3^{S_A})=\cA_3(1^{-},2^{-},3^{S_A})=0~~,~~\cA_3(1^{+},2^{-},3^{S_A})=\sqrt{2}\mu_{3A}{\spaa{2~3}\spbb{3~1}^2\over
\spaa{1~2}\spbb{1~2}\spbb{2~3}}~,~~~\nonumber\\
&&\cA_3(1^{S_A},2^{S_B},3^{+})=\delta^{AB}{\spbb{2~3}\spbb{3~1}\over\spbb{1~2}}~~,~~\cA_3(1^{S_A},2^{S_B},3^{-})=-\delta^{AB}{\spaa{2~3}\spaa{3~1}\over\spaa{1~2}}~,~~~\nonumber\\
&&\cA_3(1^{S_A},2^{S_B},3^{S_C})=\sqrt{2}\Big(\delta^{AB}\mu_{3C}{\spab{q|2|q}\over
\spab{q|3|q}}+\delta^{BC}\mu_{1A}{\spab{q|3|q}\over
\spab{q|1|q}}+\delta^{CA}\mu_{2B}{\spab{q|1|q}\over
\spab{q|2|q}}\Big)~,~~~\label{3point}\eea
where $\mu_{iA}$ is the $(4+A)$-th component of $p_i$ (or the $A$-th
component in $d$-dimension), and $|p_i\rangle, |p_i]$ are
understood to be $|p_i^\bot\rangle, |p_i^\bot]$ whenever $p_i$ is a
$D$-dimensional momentum as defined in \eref{ell-bot}. With these
inputs, we can get all tree amplitudes by BCFW recursion
relation. For simplicity, we prefer to deform a pair of momenta
which are 4-dimensional. Particularly, in this paper, we need
4-point tree amplitudes with two momenta $p_1,p_2$ in
4-dimension and the rest two $p_3,p_4$ in $D$-dimension. The
deformation pair $\spab{1^+|2}$ would suffice to compute all
required amplitudes without missing any boundary contribution. To
illustrate this, let us compute $\cA_4^{\tree}(1^+,2^+,3^+,4^-)$ for
example. Under $\spab{1|2}$ deformation, there is only one
contribution
\bea \cA_3(4^-,\widehat{1}^-,-\widehat{P}_{41}^+){1\over
s_{41}}\cA_3(\widehat{P}_{41}^-,\widehat{2}^+,3^+)~,~~~\nonumber\eea
and the deformed quantities are
\bea |\widehat{1}\rangle=|1\rangle-{P_{41}^2\over
\spab{2|P_{41}|1}}|2\rangle~~,~~|\widehat{2}]=|2]+{P_{41}^2\over
\spab{2|P_{41}|1}}|1]~~,~~|\widehat{P}_{41}\rangle
[\widehat{P}_{41}|={P_{41}|1]\langle 2|P_{41}\over
\spab{2|P_{41}|1}}~,~~~\nonumber\eea
therefore
\bea
\cA^{\tree}_4(1^+,2^+,3^+,4^-)={\spbb{1~\widehat{P}_{41}}^3\over
\spbb{\widehat{P}_{41}~4^\bot}\spbb{4^\bot~1}}{1\over
s_{14}}{\spbb{\widehat{2}~3^\bot}^3\over
\spbb{3^\bot~\widehat{P}_{41}}\spbb{\widehat{P}_{41}~\widehat{2}}}={\mu^2\over\spab{1|4|1}}{\spbb{2~1}\spaa{q~4^\bot}^2\over
\spaa{1~2}\spaa{q~3^\bot}^2}~.~~~\nonumber\eea

For reader's reference, here we list all necessary 4-point tree
amplitudes with two 4-dimensional momenta denoted as $p_1,p_2$
and two $D$-dimensional momenta denoted as
$\widehat{\ell}_1,\widehat{\ell}_2$. The polarization of $p_i$ can be
either $+$ or $-$, while for $\widehat{\ell}_i$ it can be
$+$, $-$ or $S_A$. Explicitly, without any $S$-component, we have
\bea \cA_4(1^+,2^+,\widehat{\ell}_2^{~+},\widehat{\ell}_1^{~+})&= &
0~~~,~~~\cA_4(1^+,2^+,\widehat{\ell}_2^{~+},\widehat{\ell}_1^{~-})={\mu^2
\spaa{\ell_1^\bot~q}^2\spbb{2~1}\over\spaa{\ell_2^\bot~q}^2\spaa{1~2}\spab{2|{\ell}_2|2}}~,~~~\nn
\cA_4(1^+,2^+,\widehat{\ell}_2^{~-},\widehat{\ell}_1^{~-})&=
&-{\spaa{\ell_1^\bot~\ell_2^\bot}^2\spbb{2~1}
\over\spaa{1~2}\spab{2|{\ell}_2|2}}~,~~~\label{S0-1}\eea
and
\bea \cA_4(1^+,2^-,\widehat{\ell}_2^{~+},\widehat{\ell}_1^{~+})& =
&{\mu^2~
\Spaa{q~2}^4\spbb{2~1}\over\Spaa{\ell_1^\bot~q}^2\spaa{\ell_2^\bot~q}^2\spaa{1~2}\spab{2|{\ell}_2|2}}~~,
~~\cA_4(1^+,2^-,\widehat{\ell}_2^{~-},\widehat{\ell}_1^{~-})={\mu^2~
\spaa{1~2}\spbb{1~q}^4\over\spab{2|\widehat{\ell}_2|2}\spbb{q~\ell_1^\bot}^2\spbb{q~\ell_2^\bot}^2\spbb{2~1}}~,~~~\nn
\cA_4(1^+,2^-,\widehat{\ell}_2^{~+},\widehat{\ell}_1^{~-})&= &
{\big(\spaa{1~2}\spbb{1~\ell_2^\bot}
\spbb{1~q}-\spab{2|\widehat{\ell}_2|1}\spbb{q~\ell_2^\bot}\big)\big(\spaa{\ell_1^\bot~q}\spab{2|\widehat{\ell}_1|1}
+\spaa{\ell_1^\bot~2}\spaa{q~2}\spbb{2~1}\big)\over\spaa{\ell_2^\bot~q}\spaa{1~2}
\spab{2|{\ell}_2|2}\spbb{q~\ell_1^\bot}\spbb{2~1}}~.~~~\label{S0-2}\eea
With one $S$-component we have
\bea \cA_4(1^+,2^+,\widehat{\ell}_2^{~+},\widehat{\ell}_1^{~S_A})&=
&0~~~,~~~
\cA_4(1^+,2^+,\widehat{\ell}_2^{~-},\widehat{\ell}_1^{~S_A})={\sqrt{2}\mu_A
\spaa{\ell_1^\bot~\ell_2^\bot}\spaa{\ell_2^\bot~q}\spbb{2~1}\over
\spaa{\ell_1^\bot~q}\spaa{1~2}\spab{2|{\ell}_2|2}}~,~~~\nn
\cA_4(1^+,2^-,\widehat{\ell}_2^{~+},\widehat{\ell}_1^{~S_A})&= &
{\sqrt{2}\mu_A
~\spaa{q~2}^2\big(\spaa{1~2}\spbb{1~\ell_2^\bot}\spbb{1~q}-\spab{2|\widehat{\ell}_2|1}\spbb{q~\ell_2^\bot}
\big)\over
\spaa{\ell_2^\bot~q}\spaa{1~2}\spab{q|{\ell}_1|q}\spab{2|{\ell}_2|2}}~,~~~\nn
\cA_4(1^+,2^-,\widehat{\ell}_2^{~-},\widehat{\ell}_1^{~S_A})&=
&{\sqrt{2}\mu_A~\spbb{1~q}^2\big(\spaa{\ell_2^\bot~2}\spaa{q~2}\spbb{2~1}+
\spaa{\ell_2^\bot~q}\spab{2|\widehat{\ell}_2|1}\big)\over\spab{q|{\ell}_1|q}
\spab{2|{\ell_2}|2}\spbb{q~\ell_2^\bot}\spbb{2~1}}~.~~~\label{S1}\eea
%\eea
%
With two $S$-components we have
\bea \cA_4(1^+,2^+,\widehat{\ell}_2^{~S_A},\widehat{\ell}_1^{~S_B})
& =
&-{\mu^2\spbb{2~1}\over\spaa{1~2}\spab{2|{\ell}_2|2}}\delta^{AB}~,~~~\nn
\cA_4(1^+,2^-,\widehat{\ell}_2^{~S_A},\widehat{\ell}_1^{~S_A})&=
&-{2\mu_A^2~\spaa{q~2}^2\spbb{1~q}^2\over\spab{q|{\ell}_1|q}\spab{q|{\ell}_2|q}\spab{2|\widehat{\ell}_2|2}}
-{\spab{2|\widehat{\ell}_1|1}\spab{2|\widehat{\ell}_2|1}\over
\spaa{1~2}\spab{2|\widehat{\ell}_2|2}\spbb{2~1}}~,~~~\nn
\cA_4(1^+,2^-,\widehat{\ell}_2^{~S_A},\widehat{\ell}_1^{~S_B})&=
&-{2\mu_A\mu_B~\spaa{q~2}^2\spbb{1~q}^2\over\spab{q|\widehat{\ell}_1|q}
\spab{q|{\ell}_2|q}\spab{2|{\ell}_2|2}}~,~~~A\neq
B~.~~~\label{S2}\eea
The rest of all needed amplitudes can be obtained by complex conjugation, i.e.,
$\ket{~}\leftrightarrow \bket{~}$, possibly along with the reflection identity of color-ordered amplitudes.

%%%%%%%%%%%%%%%%%%%%
\section{Details of the helicity summation for $\cA_{4}^{\oneloop}(1^+,2^+,3^+,4^-)$}
\label{HsumDetail}
%%%%%%%%%%%%%%%%%%

For generality, let us consider the $D$-dimensional loop momentum
$\WH\ell_i=(\ell,\vec{\mu})$ defined in Appendix \ref{convention}.
With respect to the diagram in Figure (\ref{Hsum}.b), we have
\bea
&&\cA_{L}(1^+,2^+,\WH\ell_R^{~+},-\WH\ell_L^{~-})\cA_{R}(\WH\ell_L^{~+},-\WH\ell_R^{~-},3^+,4^-)\nonumber\\
&=&\frac{\mu^2 \spbb{2~1} \spab{4|{\ell}_L|3} \spab{4|{\ell}_R|3}
\spab{q|{\ell}_L|q}}{\spaa{1~2} \spaa{3~4} \spbb{4~3}
\spab{q|{\ell}_R|q}(2\WH\ell_L\cdot p_1)(2\WH\ell_L\cdot
p_4)}-\frac{\mu^2 \spbb{2~1} \spbb{3~q} \spab{4|{\ell}_R|3}
\spab{q|{\ell}_L|3}}{\spaa{1~2} \spbb{4~3}
\spab{q|{\ell}_R|q}(2\WH\ell_L\cdot p_1)(2\WH\ell_L\cdot
p_4)}\nonumber\\
&&~~~~~~~~+\frac{\mu^2 \spbb{2~1} \spaa{q~4} \spbb{3~q}
\spab{q|{\ell}_L|3} \spab{4|{\ell}_R|q}}{\spaa{1~2}
\spab{q|{\ell}_R|q}^2(2\WH\ell_L\cdot p_1)(2\WH\ell_L\cdot
p_4)}-\frac{\mu^2 \spbb{2~1} \spaa{q~4} \spab{4|{\ell}_L|3}
\spab{q|{\ell}_L|q} \spab{4|{\ell}_R|q}}{\spaa{1~2} \spaa{3~4}
\spab{q|{\ell}_R|q}^2(2\WH\ell_L\cdot p_1)(2\WH\ell_L\cdot
p_4)}~.~~~\eea
For the diagram in Figure (\ref{Hsum}.c), we have
\bea
&&\cA_{L}(1^+,2^+,\WH\ell_R^{~-},-\WH\ell_L^{~+})\cA_{R}(\WH\ell_L^{~-},-\WH\ell_R^{~+},3^+,4^-)\nonumber\\
&=&\frac{\mu^2 \spbb{2~1} \spaa{q~4} \spbb{3~q} \spab{4|{\ell}_L|q}
\spab{q|{\ell}_R|3}}{\spaa{1~2}
\spab{q|{\ell}_L|q}^2(2\WH\ell_L\cdot p_1)(2\WH\ell_L\cdot
p_4)}+\frac{\mu^2 \spbb{2~1} \spaa{q~4} \spab{4|{\ell}_R|3}
\spab{4|{\ell}_L|q} \spab{q|{\ell}_R|q}}{\spaa{1~2} \spaa{3~4}
\spab{q|{\ell}_L|q}^2(2\WH\ell_L\cdot p_1)(2\WH\ell_L\cdot
p_4)}\nonumber\\
&&~~~~~~~~~~+\frac{\mu^2 \spbb{2~1} \spbb{3~q} \spab{4|{\ell}_L|3}
\spab{q|{\ell}_R|3}}{\spaa{1~2} \spbb{4~3}
\spab{q|{\ell}_L|q}(2\WH\ell_L\cdot p_1)(2\WH\ell_L\cdot
p_4)}+\frac{\mu^2 \spbb{2~1} \spab{4|{\ell}_L|3} \spab{4|{\ell}_R|3}
\spab{q|{\ell}_R|q}}{\spaa{1~2} \spaa{3~4} \spbb{4~3}
\spab{q|{\ell}_L|q}(2\WH\ell_L\cdot p_1)(2\WH\ell_L\cdot
p_4)}~.~~~\eea
For the diagram in Figure (\ref{Hsum}.d), we have
\bea
&&\cA_{L}(1^+,2^+,\WH\ell_R^{~-},-\WH\ell_L^{~-})\cA_{R}(\WH\ell_L^{~+},-\WH\ell_R^{~+},3^+,4^-)=\frac{\mu^2
\spbb{2~1} \spbb{4~3} \spaa{q~4}^4
\spbb{q|{\ell}_R|{\ell}_L|q}^2}{\spaa{1~2} \spaa{3~4}
\spab{q|{\ell}_L|q}^2 \spab{q|{\ell}_R|q}^2(2\WH\ell_L\cdot
p_1)(2\WH\ell_L\cdot p_4)}~.~~~\eea
For the diagram in Figure (\ref{Hsum}.g), we have
\bea
&&\sum_{A=1}^{\dim[\mu]}\cA_{L}(1^+,2^+,\WH\ell_R^{~-},-\WH\ell_L^{~S_A})\cA_{R}(\WH\ell_L^{~S_A},-\WH\ell_R^{~+},3^+,4^-)\nonumber\\
&=& \frac{2 (\sum_{A=1}^{\dim[\mu]}\mu_A^2) \spbb{2~1} \spaa{q~4}^2
\spbb{3~q} \spab{q|{\ell}_R|3}
\spbb{q|{\ell}_R|{\ell}_L|q}}{\spaa{1~2} \spab{q|{\ell}_L|q}^2
\spab{q|{\ell}_R|q}(2\WH\ell_L\cdot p_1)(2\WH\ell_L\cdot
p_4)}+\frac{2 (\sum_{A=1}^{\dim[\mu]}\mu_A^2) \spbb{2~1}
\spaa{q~4}^2 \spab{4|{\ell}_R|3}
\spbb{q|{\ell}_R|{\ell}_L|q}}{\spaa{1~2} \spaa{3~4}
\spab{q|{\ell}_L|q}^2(2\WH\ell_L\cdot p_1)(2\WH\ell_L\cdot
p_4)}~.~~~\eea
For the diagram in Figure (\ref{Hsum}.h), we have
\bea
&&\sum_{A=1}^{\dim[\mu]}\cA_{L}(1^+,2^+,\WH\ell_R^{~S_A},-\WH\ell_L^{~-})\cA_{R}(\WH\ell_L^{~+},-\WH\ell_R^{~S_A},3^+,4^-)\nonumber\\
&=&\frac{2 (\sum_{A=1}^{\dim[\mu]}\mu_A^2) \spbb{2~1} \spaa{q~4}^2
\spab{4|{\ell}_L|3} \spbb{q|{\ell}_R|{\ell}_L|q}}{\spaa{1~2}
\spaa{3~4} \spab{q|{\ell}_R|q}^2(2\WH\ell_L\cdot
p_1)(2\WH\ell_L\cdot p_4)}-\frac{2 (\sum_{A=1}^{\dim[\mu]}\mu_A^2)
\spbb{2~1} \spaa{q~4}^2 \spbb{3~q} \spab{q|{\ell}_L|3}
\spbb{q|{\ell}_R|{\ell}_L|q}}{\spaa{1~2} \spab{q|{\ell}_L|q}
\spab{q|{\ell}_R|q}^2(2\WH\ell_L\cdot p_1)(2\WH\ell_L\cdot
p_4)}~.~~~\eea
Finally for the diagram in Figure (\ref{Hsum}.i), we have
\bea
&&\sum_{A=1}^{\dim[\mu]}\cA_{L}(1^+,2^+,\WH\ell_R^{~S_A},-\WH\ell_L^{~S_A})\cA_{R}(\WH\ell_L^{~S_A},-\WH\ell_R^{~S_A},3^+,4^-)\nonumber\\
&=& (D-4)\frac{\mu^2 \spbb{2~1} \spab{4|{\ell}_L|3}
\spab{4|{\ell}_R|3}}{\spaa{1~2} \spaa{3~4}
\spbb{4~3}(2\WH\ell_L\cdot p_1)(2\WH\ell_L\cdot p_4)}+\frac{2
\mu^2(\sum_{A=1}^{\dim[\mu]}\mu_A^2) \spbb{2~1} \spaa{q~4}^2
\spbb{3~q}^2}{\spaa{1~2} \spab{q|{\ell}_L|q}
\spab{q|{\ell}_R|q}(2\WH\ell_L\cdot p_1)(2\WH\ell_L\cdot
p_4)}~.~~~\eea
Note that in the results above, we have explicitly written down
$\mu^2$ and $\sum_{A=1}^{\dim[\mu]}\mu_A^2$, in order to
distinguish two origins of the $\mu^2$-dependence. The factor $\mu^2$
comes directly from a single diagram, while the factor
$(\sum_{A=1}^{\dim[\mu]}\mu_A^2)=\mu^2$ comes from the summation of
helicity states $S_A$. Also, note the scalar product
$\WH\ell\cdot p_i=\ell_{{\tiny \mbox{4-dim}}}\cdot p_i$, since all
external momenta are 4-dimensional.

Although looks awful, the sum of the contributions above is
independent of the reference momentum $q$. In fact, it can be
reduced to
\bea
\sum_{h_1,h_2}\cA_{L}(1^+,2^+,\WH\ell_R^{~h_1},-\WH\ell_L^{~h_2})\cA_{R}(\WH\ell_L^{~\bar{h}_2},-\WH\ell_R^{~\bar{h}_1},3^+,4^-)=(D-2)\mu^2\frac{
\spbb{2~1} \spab{4|{\ell}_L|3}^2}{\spaa{1~2} \spaa{3~4}
\spbb{4~3}(2\WH\ell_L\cdot p_1)(2\WH\ell_L\cdot p_4)}~,~~~\eea
and this equivalence has passed the numeric test.

%%%%%%%%%%%%%%%%%%%%%%%%%%%%%%%%%%%% %%%%%%%

\bibliographystyle{JHEP}
\bibliography{Qcut}

\end{document}